\documentclass[%
 aip,
 amsmath,amssymb,
 reprint,%
]{revtex4-1}
\usepackage{graphicx}% Include figure files
\usepackage{xcolor, soul}
\usepackage[breaklinks=true,colorlinks=true]{hyperref}
\usepackage{dcolumn}% Align table columns on decimal point
\usepackage{bm}% bold math
\usepackage[utf8]{inputenc}
\usepackage[T1]{fontenc}
\usepackage{mathptmx}

\newcommand{\beginsupplement}{%
        \setcounter{table}{0}
        \renewcommand{\thetable}{S\arabic{table}}%
        \setcounter{figure}{0}
        \renewcommand{\thefigure}{S\arabic{figure}}
        \setcounter{equation}{0}
        \renewcommand{\theequation}{S\arabic{equation}}%
     }

\begin{document}

\preprint{AIP/123-QED}

%\title{Fabricating Hybrid Magnon-Polariton Devices using Plasma Focused Ion Beam Processing}
\title{Strong magnon-photon coupling with chip-integrated YIG in the zero-temperature limit}%\M{is 'zero-temperature limit' essential? More than the nanomanipulator?}}
\author{Paul G.~Baity}% \M{full name? Contact email}}
\email{paul.baity@glasgow.ac.uk}
\affiliation{James Watt School of Engineering, Electronics \& Nanoscale Engineering Division, University of Glasgow, Glasgow G12 8QQ, United Kingdom}%Lines break automatically or can be forced with \\
\author{Dmytro A. Bozhko}%
\affiliation{Center for Magnetism and Magnetic Materials, Department of Physics and Energy Science, University of Colorado Colorado Springs, Colorado Springs, Colorado 80918, USA}
\author{Rair Mac\^edo}
\affiliation{James Watt School of Engineering, Electronics \& Nanoscale Engineering Division, University of Glasgow, Glasgow G12 8QQ, United Kingdom}
\author{William Smith}
\affiliation{SUPA, School of Physics and Astronomy, University of Glasgow, Glasgow G12 8QQ, United Kingdom}
\author{Rory C. Holland}
\affiliation{James Watt School of Engineering, Electronics \& Nanoscale Engineering Division, University of Glasgow, Glasgow G12 8QQ, United Kingdom}
\author{Sergey Danilin}
\affiliation{James Watt School of Engineering, Electronics \& Nanoscale Engineering Division, University of Glasgow, Glasgow G12 8QQ, United Kingdom}
\author{Valentino Seferai}
\affiliation{James Watt School of Engineering, Electronics \& Nanoscale Engineering Division, University of Glasgow, Glasgow G12 8QQ, United Kingdom}
\author{Jo\~{a}o Barbosa}
\affiliation{James Watt School of Engineering, Electronics \& Nanoscale Engineering Division, University of Glasgow, Glasgow G12 8QQ, United Kingdom}
\author{Renju R. Peroor}%
\affiliation{Center for Magnetism and Magnetic Materials, Department of Physics and Energy Science, University of Colorado Colorado Springs, Colorado Springs, Colorado 80918, USA}
\author{Sara Goldman}%
\affiliation{Center for Magnetism and Magnetic Materials, Department of Physics and Energy Science, University of Colorado Colorado Springs, Colorado Springs, Colorado 80918, USA}
\author{Umberto Nasti}
\affiliation{James Watt School of Engineering, Electronics \& Nanoscale Engineering Division, University of Glasgow, Glasgow G12 8QQ, United Kingdom}
\affiliation{Current affiliation: School of Engineering \& Physical Sciences, Heriot-Watt University, Edinburgh EH14 4AS, United Kingdom
}
\author{Jharna Paul}
\affiliation{James Watt School of Engineering, Electronics \& Nanoscale Engineering Division, University of Glasgow, Glasgow G12 8QQ, United Kingdom}
\author{Robert H. Hadfield}
\affiliation{James Watt School of Engineering, Electronics \& Nanoscale Engineering Division, University of Glasgow, Glasgow G12 8QQ, United Kingdom}
\author{Stephen McVitie}
\affiliation{SUPA, School of Physics and Astronomy, University of Glasgow, Glasgow G12 8QQ, United Kingdom}
\author{Martin Weides}
\affiliation{James Watt School of Engineering, Electronics \& Nanoscale Engineering Division, University of Glasgow, Glasgow G12 8QQ, United Kingdom}

\date{\today}% It is always \today, today,

\begin{abstract}
The cross-integration of spin-wave and superconducting technologies is a promising method for creating novel hybrid devices for future information processing technologies to store, manipulate, or convert data in both classical and quantum regimes. Hybrid magnon-polariton systems have been widely studied using bulk Yttrium Iron Garnet (Y$_{3}$Fe$_{5}$O$_{12}$, YIG) and three-dimensional microwave photon cavities. However, limitations in YIG growth have thus far prevented its incorporation into CMOS compatible technology such as high quality factor superconducting quantum technology. To overcome this impediment, we have used Plasma Focused Ion Beam (PFIB) technology---taking advantage of precision placement down to the micron-scale---to integrate YIG with superconducting microwave devices. Ferromagnetic resonance has been measured at milliKelvin temperatures on PFIB-processed YIG samples using planar microwave circuits. Furthermore, we demonstrate strong coupling between superconducting resonator and YIG ferromagnetic resonance modes by maintaining reasonably low loss while reducing the system down to the micron scale. This achievement of strong coupling on-chip is a crucial step toward fabrication of functional hybrid quantum devices that advantage from spin-wave and superconducting components.
\end{abstract}

\maketitle

\begin{figure}[t!]%{L}{0.5\textwidth}
%\centering
\includegraphics[scale=0.45]{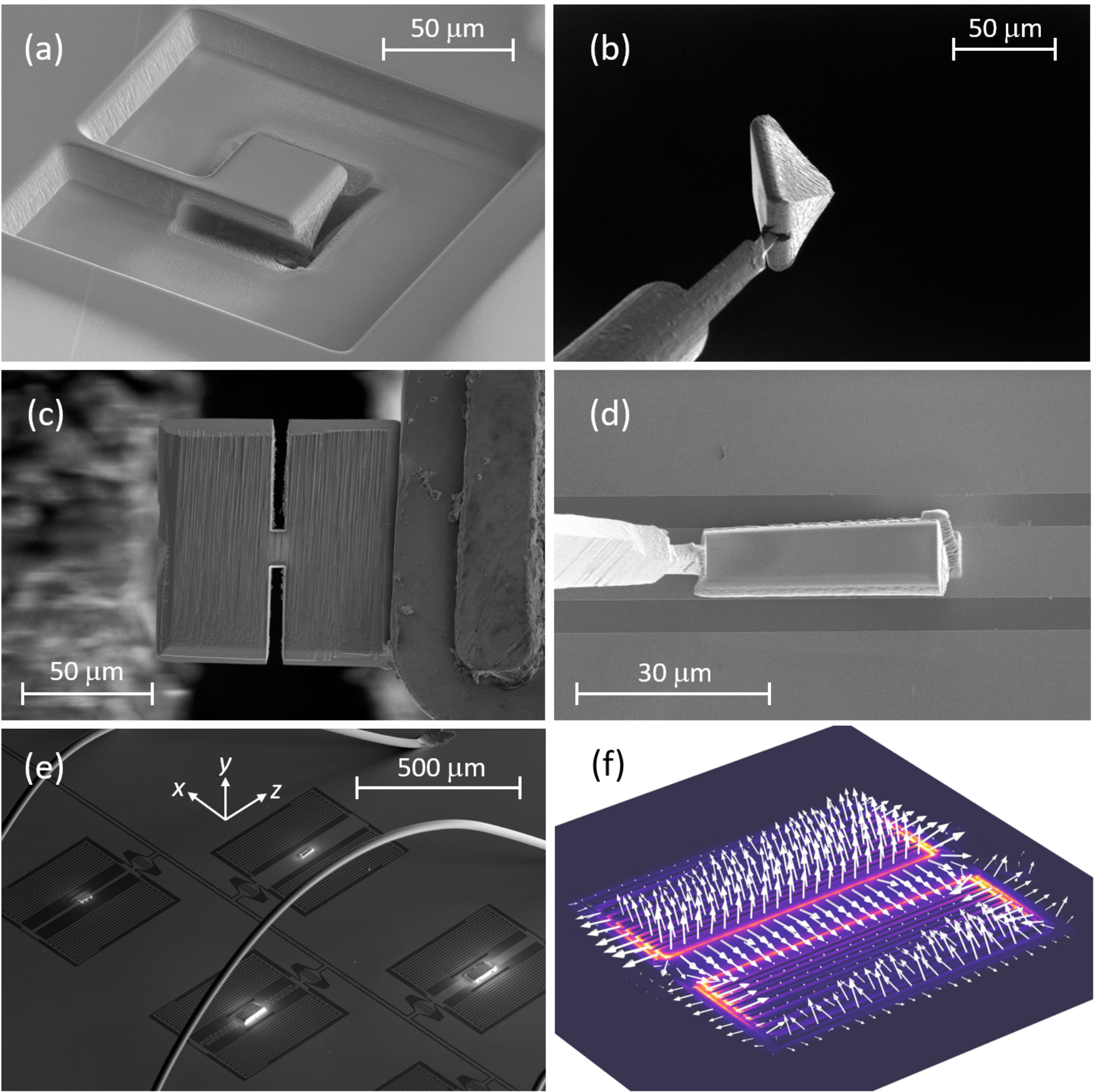}
\caption{SEM images showing the PFIB processing steps for YIG. (a) Section of a YIG film that has been milled out using a Xe ion beam. An upside down pyramid structure with an approximate $50 \times 50$~$\mu$m$^2$ base is milled from the surface of the YIG film. (b) The pyramid is attached to a nanomanipulator via Pt deposition on the manipulator tip. Afterward, the connecting bridge (seen in (a)) is milled away and the pyramid is extracted from the film surface as shown. (c) The extracted pyramid from (b) is attached to a metal finger (right object) using Pt deposition. From this position, the pyramid is milled flat into a cuboid shape. The cuboid can be cut into smaller sections (e.g.~$50\times 10\times 5$~$\mu$m$^3$). (d) The samples are then placed onto the NbN microwave devices using the nanomanipulator. The YIG samples are affixed to the device surfaces by precision Pt deposition. The YIG is milled free from the nanomanipulator by milling briefly near the area connecting the nanomanipulator and YIG sample. (e) An SEM image of NbN lumped element resonators with YIG placed onto the central inductive line where the magnetic antinode is. The YIG samples are oriented with respect to the shown axes. (f) Simulation of the NbN resonator with zero-field frequency $\omega_{p0}/2\pi = 12.517$~GHz. The field profiles were determined for the bare resonator without YIG and simulated using COMSOL Multiphysics\textsuperscript{\textregistered} software. \label{fig1}}
\end{figure}

In recent years, there has been large investment into research on cavity magnon-polariton systems with strongly coupled spin-wave and photonic components\cite{Huebl2013, Tabuchi2014, Zhang2014, Zhang2015, Morris2017, Boventer2018, Boventer2020, Trempler2020}. This research is driven in part by the promise of creating new hybrid devices using spin-wave and other technologies, such as superconducting quantum technology\cite{Tabuchi2015}. Indeed, hybrid ferromagnetic-superconducting systems have been realized using both a 3D electromagnetic cavity\cite{Tabuchi2015} and planarized 2D resonators\cite{Li2019}. This latter work made a large step toward full integration of spin-wave and superconducting technologies, however, the large magnon line-widths of ferromagnetic metals, such as permalloy (Ni$_{80}$Fe$_{20}$), pose deleterious constraints on quantum coherent applications. The ferrimagnetic insulator yttrium iron garnet (Y$_{3}$Fe$_{5}$O$_{12}$, YIG) on the other hand, has a spin-wave damping which is orders of magnitude lower that metallic ferromagnets as well as much longer magnon lifetimes\cite{Cherepanov1993, Kosen2019, Mihalceanu2018}. Therefore, this material has been the forefront candidate for several applications in the field of spin-wave technology (aka magnonics\cite{Kruglyak2010, Serga2010, Lenk2011, Fischer2017, Frey2020, Prokopenko2019}). However, despite its exceptional spin-wave qualities, YIG poses issues of its own, specifically its difficult growth and incompatibility standard processing techniques. The growth of high-quality YIG typically requires specialized substrates such as gadolinium gallium garnet (Gd$_{3}$Ga$_{5}$O$_{12}$, GGG), making it incompatible with CMOS processing techniques\cite{Lenk2011}. Here we present a method for integrating high-quality YIG with superconducting quantum technologies using plasma focused ion beam (PFIB) technologies. Our primary result is the achievement of strong coupling on chip devices comprising a superconducting resonator and a YIG sample  at sub-Kelvin temperatures.

%\section{Device Fabrication}

\begin{table*}[t]
 \begin{center}
 \begin{tabular}{ | c | l | l | l | l | l | l |}
 \hline
  Sample & $x$ [$\mu$m] & $y$ [$\mu$m] & $z$ [$\mu$m] & $D_x$ & $D_y$ & $D_z$\\ 
  \hline
  \# 1 & $50\pm6$ & $15\pm1$ &  $45.8\pm0.3$ & $0.187 \pm 0.004$  & $0.61 \pm 0.01$ & $0.204 \pm 0.002$\\
  \hline
  \# 2 &  $10.8\pm0.5$ & $6.1\pm0.3$ & $36.3\pm0.4$ & $0.335 \pm 0.004$ & $0.570 \pm 0.007$ & $0.0958 \pm 0.0003$\\ 
  \hline
  \# 3 & $12.5\pm0.5$ & $6.5\pm0.5$ & $47.5\pm0.5$ & $0.326 \pm 0.005$ & $0.59 \pm 0.01$ & $0.0818 \pm 0.0003$\\
  \hline
%\caption{caption}
 \end{tabular}
 \end{center}
 \caption{
 Sample dimensions and their calculated demagnetisation factors. Errors on sample dimensions correspond to the deviations from the cuboid shape across each sample. The demagnetization factors were calculated from sample dimensions using equations in Ref.~\onlinecite{Aharoni1998}. Errors on the demagnetization factors reflect the shape errors. For the sample dimensions, the $z$ direction is oriented parallel to the applied field, along the device transmission line length (see insets of Figs.~\ref{fig2}(a) and (b)). The $x$ and $y$ directions are oriented to the respective in-plane and out-of-plane perpendiculars to $z$ axis. \label{DemagTable}}
\end{table*}

The hybrid superconducting and ferromagnetic devices in this study were fabricated in two stages: the fabrication of planar superconducting devices and the integration of PFIB-processed YIG onto those devices. Superconducting planar devices were fabricated from 100~nm thick niobium nitrate (NbN) films. The films were deposited onto intrinsic silicon substrates at room temperature using reactive sputtering of Nb in a mixture of N$_2$ and Ar gas flow in a similar fashion to other studies\cite{Banerjee2018}. NbN material was chosen for its high critical field values, which ensures that the devices remain superconducting and operable under applied field.
The superconducting transition temperature was found to be $\sim 11$~K using four-probe resistivity measurements. Planar devices were fabricated from the NbN films using photo- and electron-beam lithography and reactive ion etching methods.

YIG samples were processed from a 100 $\mu$m thick film, grown on GGG, using Xe plasma focused ion beam. Processing steps for the YIG samples are shown in Fig.~\ref{fig1}(a-d). First, a large section of the film is first milled out using a wide beam column. The sample is then tilted by 52$^\circ$ relative to the Xe beam and a narrower column beam is used to mill beneath the desired section of the film, creating an inverted pyramid shape of YIG (Fig.~\ref{fig1}(a)). Before the sample is fully removed from the surface of the film, the finger of a nanomanipulator approaches the material's edge, where Pt is deposited, connecting the sample to the finger. The pyramid-shaped YIG can then be extracted on the nanomanipulator, as shown in Fig.~\ref{fig1}(b). After the sample is fixed to the nanomanipulator, it can be fully removed from the surface and moved to a stage, where finer milling can be performed to achieve the desired shape (Fig.~\ref{fig1}(c)). Afterwards, the sample is removed again by the nanomanipulator and placed on top of prefabricated superconducting microwave devices (Fig.~\ref{fig1}(d)). Currents for Xe milling ranged from 1.8 to 180 nA. For the three samples (\#1-3) reported in this work, pieces of YIG were shaped into rough rectangular cuboid shapes with dimensions ranging from 5 to 50~$\mu$m.  The measured dimensions for each sample are listed in Table~\ref{DemagTable}. These pieces were placed, with micron precision, on top of NbN microwave devices for spectral measurement. The YIG samples were fixed to these devices using precise Pt deposition. When necessary, ion milling was used to remove excess Pt that could short the microwave devices (an example of this can be seen in Fig.~\ref{fig1}(e) as dark lines in the gaps between the central inductive lines and interdigitated capacitors of the superconducting resonator devices).

\begin{figure}
\includegraphics[scale=0.9]{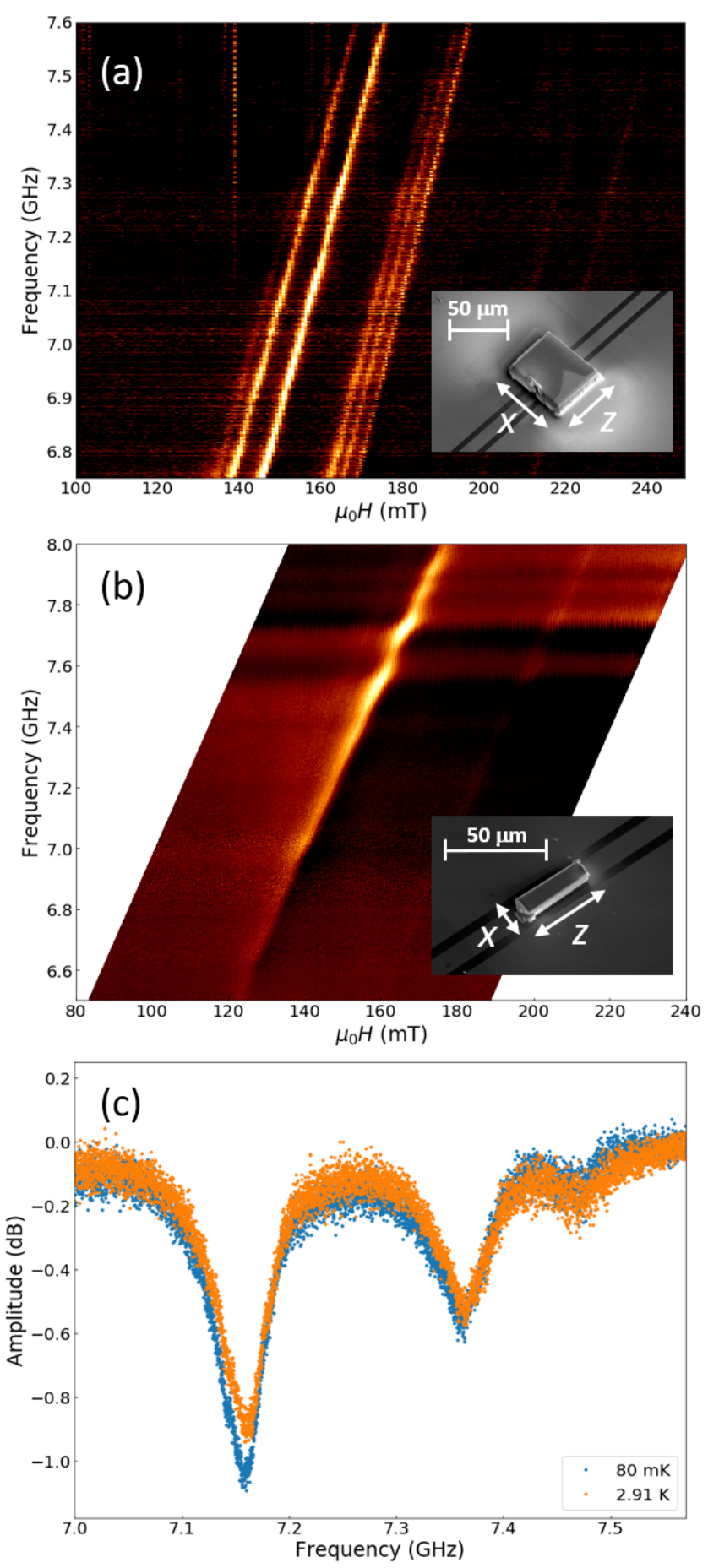}
\caption{(a) Spectral measurement of sample 1 on top of a 10 $\mu$m wide NbN transmission line at 3 K. Inset: SEM image of sample 1. The $x$ and $z$ axes of the sample are labelled. The $y$ axis (unlabelled) is oriented perpendicular to the NbN film surface. (b) Measurement of sample 2 on top of a 10 $\mu$m wide NbN transmission line at 2.9~K. Inset: SEM image of the sample 2 with labelled axes. (c) FMR resonances of sample 1 at $\mu_0H=160$~mT. The FMR linewidths show no significant change from 2.9~K to 80~mK.} \label{fig2}
\end{figure}

Ferromagnetic resonance spectra were measured on YIG samples 1 and 2. The samples were placed on top of NbN coplanar waveguide transmission lines with a 10~$\mu$m width, 6~$\mu$m gap geometry (see insets of Fig.~\ref{fig2}(a) and (b)). The devices were then placed inside Cu boxes for microwave measurement and mounted inside a superconducting solenoid within an adiabatic demagnetization refrigerator (see supplement Fig.~S1(a) for a diagram of the measurement setup). The solenoid field was calibrated using an electron spin resonance measurement of a known material\cite{Voesch2015} (see supplement Fig.~S1(b) and text). Microwave transmission, $S_{21}$, was measured by a vector network analyzer (VNA) sending signals to ports on the NbN transmission line. The VNA signals were attenuated by 20~dB at 70~K, 4~K, and 0.5~K stages before reaching the transmission line. At the 4~K stage, a high-electron-mobility transistor (HEMT) provided 40~dB amplification to signals after passing through the chip. An isolator placed before the HEMT prevented contamination of transmission signals from spurious reflections back to the transmission line. For some measurements, additional 45~dB amplification was provided by a room temperature amplifier. VNA excitations ranged from -30 to 0~dBm, and no power dependence was observed in this range for any samples. Ferromagnetic resonance (FMR) was then measured at 3.2~K and 80~mK by applying a static magnetic field parallel to the transmission line via the superconducting solenoid. For all measurements presented, the signals are normalized with respect to the microwave background signal.

As shown in Fig.~\ref{fig2}(a-c), the FMR lines appear in the spectrum as absorbing resonances whose frequency increases with applied field. The spectra have a strong dependence on sample shape and size. For example, sample 1, with the largest volume, exhibits several spin-wave modes (Fig.~\ref{fig2}(a)). We associate these additional resonances with a combination of magnetostatic surface spin waves and forward and backward volume magnetostatic spin waves\cite{Serga2010}. However, identification of the modes is difficult due to the imperfect cuboid sample shape and potential effects on the spin waves from the coplanar waveguide geometry of the superconducting transmission line\cite{Golovchanskiy2019}. On the other hand, if the width and thickness (i.e. $x$ and $y$ dimensions) of the YIG are reduced, such as for sample 2, then modes become more isolate, leaving only one dominant mode in the spectrum (Fig.~\ref{fig2}(b)). We associate this resonance in the spectrum of sample 2 with the Kittel mode\cite{Kittel1948}. Our main focus is on the Kittel mode, whose resonance frequency is described according to the Kittel equation\cite{Kittel1948}
\begin{equation}
\omega_{FMR}=\gamma \mu_0 \sqrt{\left[H_{eff} + (D_x-D_z) M_s\right]\cdot \left[H_{eff} + (D_y-D_z) M_s\right]}.\label{fmreq}
\end{equation} 
Here $\gamma$ is the gyromagnetic ratio, $\mu_0$ is the permeability of free space, $M_s$ is the saturation magnetization, and $D_x$, $D_y$, and $D_z$ are demagnetization factors, which account for the effect of the sample shape. $H_{eff} = H + H_k$ is the effective field consisting of the  external field $H$ produced by the superconducting solenoid applied along the $z$ direction and the field $H_k$ associated with the YIG magnetocrystalline anisotropy. The field-dependent resonance frequency for sample 2 can be fit to this equation to determine the saturation magnetization of the PFIB-processed YIG. First, however, the demagnetization factors are calculated for the cuboid shape of the sample by using equations supplied in Ref.~\onlinecite{Aharoni1998}. Using these factors, which are listed in Table \ref{DemagTable}, the saturation magnetization is determined to be $234 \pm 4$~mT in reasonable agreement with other measurements on YIG near zero temperature (see supplement Fig.~S3(b) and Refs.~\onlinecite{Boventer2018, Mihalceanu2018}). The magnetocrystalline anisotropy, which depends on $M_s$, is determined to be $\mu_0H_k\approx 30$~mT and will be discussed in more detail below. However, due to the weakness of the FMR signal, a determination of the coupling strength is not possible, and therefore an extraction of the magnon linewidths from the FMR data is unreliable. Nevertheless, from fits of the resonance signals of samples 1-3, we can determine a linewidth range of 15-40~MHz. Furthermore, as shown in Fig.~\ref{fig2}(c), we find that the FMR modes show little to no change between $2.9$~K and $80$~mK, indicating saturation of the linewidth in the zero-temperature limit.

The main result of this work is the achievement of strong coupling between the FMR modes of the PFIB processed YIG and the photonic mode of a superconducting resonator. A scanning electron microscopy (SEM) image of the lumped-element resonator devices, which consist of LC circuits comprised of interdigitated capacitors connected by an inductive line, is shown in Fig.~\ref{fig1}(e). Using the PFIB processing techniques, YIG was placed on the center of the inductive line of the shown resonators. At this position, the field profile derived from microwave simulations, such as shown in Fig.~\ref{fig1}(f), confirms that the YIG is exposed to a uniform magnetic excitation field parallel with the chip surface and directed across the line. This ensures the strongest coupling of both resonant systems as well as that the FMR condition is met by having the excitation field perpendicular to the externally applied field from the solenoid, which is applied parallel to the inductive line of the resonator. Of the devices shown in Fig.~\ref{fig1}(e), we focus on the YIG-resonator device which exhibited well-defined signatures of strong coupling; the spectra of other select devices on the same chip are shown in Fig.~S2 of the supplement. The device of interest, which is called sample 3, has a zero-field resonance frequency of $\omega_{p0}/2\pi = 12.517$~GHz with a full-width at half-maximum linewidth of $\Gamma/\pi = 9.2$~MHz. At zero field, the loaded quality factor is $1250$ after placement of the YIG. The resonance exhibits Fano-like\cite{Fano1961,Limonov2017} behavior (see Fig.~S3(a) in the supplement), indicating the presence of interference. Such interference could be due to imperfections created by the placement of YIG or coupling of the resonator to stray resonances in the background transmission.

\begin{figure}
\includegraphics[scale=0.87]{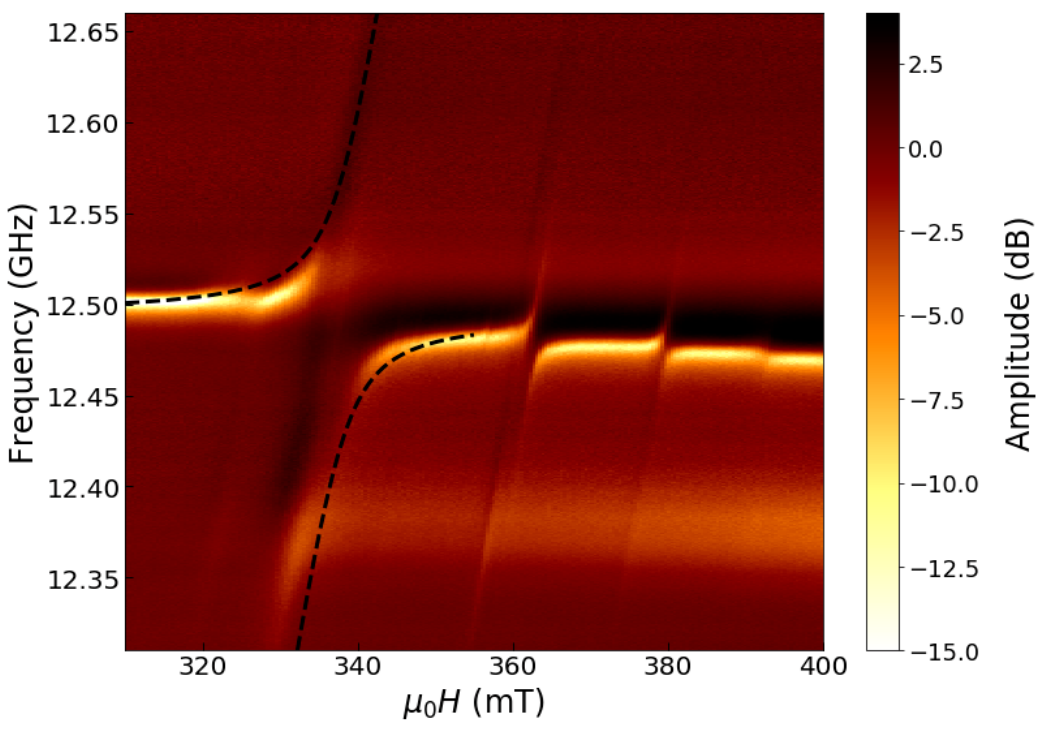}
\caption{Spectral measurement of the YIG-resonator device (sample 3) as a function of external applied field $H$. Several avoided crossings occur as each YIG FMR mode crosses the NbN resonator frequency. The black dashed lines represent fits of the branches of the largest anticrossing based on Eq.~\ref{coupledmodes}. The gap between the two branches corresponds to twice the effective coupling. For the largest anticrossing, the effective coupling is estimated to be $g/2\pi = 63\pm5$~MHz.\label{fig3}}
\end{figure}

Spectral measurements of the device were performed using the same methods as used to measure FMR of samples 1 and 2. As shown in Fig.~\ref{fig3}, when field is applied to the chip, the YIG FMR frequency increases, crossing the resonator frequency. When the two resonances coincide, coupling lifts the degeneracy of the system, resulting in an avoided crossing of modes in the spectrum. For our device, we observe several avoided crossings corresponding to coupling between the resonator photons and different spin-wave modes. The frequency separation between the branches of the avoided crossings is related to the strength of coupling, and this can be described generally by an equation of the form\cite{Huebl2013,RairsPaper}
\begin{equation}
    \omega_{\pm} = \frac{1}{2}\left[\omega_p + \omega_{FMR} \pm \sqrt{(\omega_p-\omega_{FMR})^2 + 4g^2}\right].
    \label{coupledmodes}
\end{equation}
Here, $\omega_p$ is the resonator photon frequency and $g$ is the coupling constant. Figure~\ref{fig3} shows fits of the branches of the largest avoided crossing using Eq.~\ref{coupledmodes}. We use a least-squares fitting procedure to independently fit the two branches. Saturation magnetization $M_s$ and $g$ are used as fit parameters while $\omega_p/2\pi$ is fixed at 12.494~GHz. Although the resonator frequency has a field dependence due to the proliferation of vortices in the superconductor,\cite{Bothner2012,Li2019} the field range for the avoided crossing is small enough that an adequate fit can be accomplished without including this effect. From fits of the two branches result, we find that $\mu_0 M_s=205 \pm 5$~mT and $g/2\pi = 63 \pm 5$~MHz
for the largest avoided crossing, where the errors reflect the disagreement between the two independent fits of each branch.

The coupling constant itself has been quoted as a function of the spin density, $\rho$, taking the form\cite{Huebl2013}
\begin{equation}
g = \frac{\gamma}{2}\sqrt{\frac{\mu_0\hbar\omega_p \rho V_m}{2V_p}},
\label{couplingequation}
\end{equation}
where $V_m$ is the volume of the magnetic sample and $V_p$ is the resonator's mode volume.
Here, we take the spin density to be $\rho=2 \times 10^{22}$~cm$^{-3}$ (appropriate for good-quality YIG\cite{Huebl2013}) and $V_p=1.59 \times 10^{-5}$~cm$^3$ (estimated using the finite-element simulations package COMSOL Multiphysics\textsuperscript{\textregistered}). 
Substituting these values into Eq.~\ref{couplingequation}, we find $g=70 \pm 3$~MHz which is in reasonable agreement with the fits of the experimental data. The coupling $g$ can also be theoretically derived using electromagnetic perturbation theory (see Ref.~\onlinecite{RairsPaper} and also the supplement). In this case, $g$ is defined in terms of the magnetization rather than spin density. Using $\mu_0 M_s=205$~mT, the coupling is predicted to be $g=68\pm3$~MHz, in close agreement with the value based on an assumed spin density.

Without the inclusion of magnetocrystalline anisotropy, the saturation magnetization of the PFIB-processed YIG shows consistent values of $\mu_0 M_s \sim300$~mT for samples 2 and 3. Since the Kittel mode cannot be reliably identified for sample 1, we cannot provide a proper estimate of $M_s$. Nevertheless, in comparison to other studies on YIG in the sub-Kelvin regime\cite{Boventer2018, Kosen2019, MaierFlaig2017}, the extracted values from our samples would initially seem much higher than previously reported. While demagnetization factors are taken into account, the high $M_s$ values imply that other effects such as magnetocrystalline anisotropy should be considered. At 4.2~K, the first and secondary cubic magnetocrystalline anisostropy constants\cite{SpinWavesBook} for YIG are $K_1 = -2480$~J/m$^3$ and $K_2 = -118$~J/m$^3$. Although the samples are cuboid in shape, the internal anisotropy fields can be approximated by a relation for a (111)-oriented 2D film\cite{Ross2011,Lee2016}. While the exact calculation of these fields depends on the in-plane crystallographic orientation, which is lost during the PFIB processing, an order of magnitude approximation is $\mu_0 H_k \approx -2K_1/M_s - 2K_2/M_s$. By including this $M_s$-dependent field term in Eq.~\ref{fmreq}, we determine $\mu_{0}H_k \approx 30$~mT and saturation magnetization values $\mu_0 M_s = 234 \pm 4$~mT (sample 2) and $205 \pm 5$~mT (sample 3).
These values of $M_s$ are more in line with the saturation magnetization $\mu_0M_s=240\pm1$~mT found for an unstructured YIG film (see Fig.~S3(b) in the supplement) and is also in the range of previously reported values for pure YIG at $4$~K \cite{Mihalceanu2018,Algra1982}. Unfortunately, the PFIB-processed samples' magnetic moments were below the sensitivity of the available SQUID magnetometer. Previous studies indicate that ion implantation can significantly decrease the  magnetization of YIG \cite{Algra1982}, so without other means of investigation (e.g. Brillouin light scattering \cite{Agrawal2013}), we cannot make a strong argument about presence of these effects in our samples. However, a thorough study of anisotropies and magnetization behaviors after PFIB fabrication is beyond the scope of this article and is a subject for a future work.

The experimental and theoretical values of $g$ for sample 3 show relatively good agreement with only $\sim10\%$ difference. Since shape variation is already accounted for in the uncertainty on the theoretical value, it cannot completely account for the difference. Instead, the $10\%$ difference indicates that either $V_p$ is slightly underestimated in the simulation or the effective moment density $\rho$ is reduced due to damage from the PFIB processing. Although neither $V_p$ nor $\rho$ can be measured directly, there is some hint when comparing the spin-density and electromagnetic perturbation theory based predictions of $g$. Since the electromagnetic perturbation theory uses $M_s$, which is determined from fitting, the general agreement between spin-density and electromagnetic perturbation predictions of $g$ indicates that $\rho$ is close to the expected value for ideal YIG, even though the determined values of $M_s$ are different for samples 2 and 3. 
Therefore, it seems probable that surface damage from PFIB processing does not strongly affect the effective spin density and that a misestimation of $V_p$ is responsible for the disagreement between experimental and theoretical determinations of $g$. Furthermore, this also implies that the difference between $M_s$ for samples 2 and 3 may depend on other factors such as $H_k$ and the crystalline orientation, which is unaccounted for. 

Based on the experimental determination of $g$ and assumed spin-density for the first avoided crossing, the single-spin coupling constant can be determined as $g_0/2\pi = g/2\pi\sqrt{N} = 7.2$~Hz, where $N = \rho V_m$.
In addition, sample 3 exhibited at least two more at higher fields, indicating the presence of strong magnon-photon coupling to higher order spin-wave modes. Indeed, the cooperativity\cite{Boventer2018} $C=g^2/\Gamma\kappa$ can be calculated using the coupling $g$ and resonator and magnon linewidths ($\Gamma$ and $\kappa$, respectively) to confirm the strong coupling regime for all three avoided crossings. 
%While the respective linewidths can be difficult to extract precisely from the data, we note that the resonance signals have an approximately 10~MHz linewidth. Therefore, for the first avoided crossing at $\mu_0H=337$~mT, even in the worst-case scenario where $\Gamma/2\pi=\kappa/2\pi=10$~MHz, $C\approx40$, confirming the sample is firmly in the strong coupling regime where $C\gg1$. 
While the respective linewidths can be difficult to extract precisely from the data, they are estimated as $\kappa = 40 \pm 5$~MHz and $\Gamma = 6.4 \pm 0.8$~MHz from fits of $S_{21}$ at $\mu_0H = 337.5$~mT (see supplement). Therefore the cooperativity is approximately $C\approx15$ for the first avoided crossing, confirming the sample is firmly in the strong coupling regime where $C\gg1$. Furthermore, using the resonator linewidth information and following the method described in the supplement, the average number of magnons $\langle n_m \rangle$ and photons $\langle n_p \rangle$ in the system can be estimated as $\langle n_p \rangle = \langle n_m \rangle = (6 \pm 2)\times 10^3$.

While the first avoided crossing has a coupling $g/2\pi = 63\pm 5$~MHz, each subsequent avoided crossing shrinks in gap size and coupling with $g/2\pi=18\pm 2$~MHz for the second crossing at $\mu_0H=362$~mT and $g/2\pi = 10\pm 10$~MHz for the third at $\mu_0H = 380$~mT. Though the exact nomenclature of these higher-order width or thickness spin-wave modes is undetermined, in the context of Eq.~\ref{couplingequation}, the smaller coupling rates imply the effective $\rho$ is heavily reduced for these modes (i.e. fewer spins participate in coupling) in comparison to the Kittel mode, in which the spins precess together as a singular macrospin. The reduced coupling rates also imply a steadily reducing cooperativity, and therefore it can be concluded that future applications relying on higher order spin-wave modes will need to achieve even larger coupling rates to remain in the strong coupling regime. Nevertheless, the study here shows that such coupling rates may be possible using PFIB-processed YIG.

Therefore, we have established a general method for creating on-chip hybrid devices with high-quality spin-wave and superconducting components using PFIB technology. Using these methods, we fabricated several devices and demonstrated strong coupling between YIG FMR modes and a superconducting resonator mode. This serves as a crucial step toward creating hybrid quantum devices with spin-wave and superconducting components. The techniques used here can be used to create any number of photon-magnon devices and circuits such as hybrid logical gates\cite{Kostylev2005}, low-temperature multiplexers\cite{Vogt2014}, or spin-based quantum memory.

\noindent\textbf{Supplementary Material}

Further experimental, device characterization, and analysis details, along with corresponding figures, can be found in the supplementary material.

\noindent\textbf{Acknowledgements}

Authors would like to thank I.~I.~Syvorotka for the fruitful discussions. This work was supported by the European Research Council (ERC) under the Grant Agreement 648011. R.~Mac\^edo acknowledges support from the Leverhulme Trust, and the University of Glasgow through LKAS funds. R.~C.~Holland was supported by the Engineering and Physical Sciences Research Council (EPSRC) through the Vacation Internships Scheme. S.~McVitie and W. Smith acknowledge support from EPSRC grant EP/M024423/1. RHH acknowledges support through ERC grant 648604 and a Royal Society Leverhulme Trust Senior Research Fellowship. The data that support the findings of this study are available from the corresponding author upon reasonable request.

\bibliography{bib}

\begin{thebibliography}{38}%
\makeatletter
\providecommand \@ifxundefined [1]{%
 \@ifx{#1\undefined}
}%
\providecommand \@ifnum [1]{%
 \ifnum #1\expandafter \@firstoftwo
 \else \expandafter \@secondoftwo
 \fi
}%
\providecommand \@ifx [1]{%
 \ifx #1\expandafter \@firstoftwo
 \else \expandafter \@secondoftwo
 \fi
}%
\providecommand \natexlab [1]{#1}%
\providecommand \enquote  [1]{``#1''}%
\providecommand \bibnamefont  [1]{#1}%
\providecommand \bibfnamefont [1]{#1}%
\providecommand \citenamefont [1]{#1}%
\providecommand \href@noop [0]{\@secondoftwo}%
\providecommand \href [0]{\begingroup \@sanitize@url \@href}%
\providecommand \@href[1]{\@@startlink{#1}\@@href}%
\providecommand \@@href[1]{\endgroup#1\@@endlink}%
\providecommand \@sanitize@url [0]{\catcode `\\12\catcode `\$12\catcode
  `\&12\catcode `\#12\catcode `\^12\catcode `\_12\catcode `\%12\relax}%
\providecommand \@@startlink[1]{}%
\providecommand \@@endlink[0]{}%
\providecommand \url  [0]{\begingroup\@sanitize@url \@url }%
\providecommand \@url [1]{\endgroup\@href {#1}{\urlprefix }}%
\providecommand \urlprefix  [0]{URL }%
\providecommand \Eprint [0]{\href }%
\providecommand \doibase [0]{http://dx.doi.org/}%
\providecommand \selectlanguage [0]{\@gobble}%
\providecommand \bibinfo  [0]{\@secondoftwo}%
\providecommand \bibfield  [0]{\@secondoftwo}%
\providecommand \translation [1]{[#1]}%
\providecommand \BibitemOpen [0]{}%
\providecommand \bibitemStop [0]{}%
\providecommand \bibitemNoStop [0]{.\EOS\space}%
\providecommand \EOS [0]{\spacefactor3000\relax}%
\providecommand \BibitemShut  [1]{\csname bibitem#1\endcsname}%
\let\auto@bib@innerbib\@empty
%</preamble>
\bibitem [{\citenamefont {Huebl}\ \emph {et~al.}(2013)\citenamefont {Huebl},
  \citenamefont {Zollitsch}, \citenamefont {Lotze}, \citenamefont {Hocke},
  \citenamefont {Greifenstein}, \citenamefont {Marx}, \citenamefont {Gross},\
  and\ \citenamefont {Goennenwein}}]{Huebl2013}%
  \BibitemOpen
  \bibfield  {author} {\bibinfo {author} {\bibfnamefont {H.}~\bibnamefont
  {Huebl}}, \bibinfo {author} {\bibfnamefont {C.~W.}\ \bibnamefont
  {Zollitsch}}, \bibinfo {author} {\bibfnamefont {J.}~\bibnamefont {Lotze}},
  \bibinfo {author} {\bibfnamefont {F.}~\bibnamefont {Hocke}}, \bibinfo
  {author} {\bibfnamefont {M.}~\bibnamefont {Greifenstein}}, \bibinfo {author}
  {\bibfnamefont {A.}~\bibnamefont {Marx}}, \bibinfo {author} {\bibfnamefont
  {R.}~\bibnamefont {Gross}}, \ and\ \bibinfo {author} {\bibfnamefont
  {S.~T.~B.}\ \bibnamefont {Goennenwein}},\ }\bibfield  {title} {\enquote
  {\bibinfo {title} {High cooperativity in coupled microwave resonator
  ferrimagnetic insulator hybrids},}\ }\href {\doibase
  10.1103/PhysRevLett.111.127003} {\bibfield  {journal} {\bibinfo  {journal}
  {Phys. Rev. Lett.}\ }\textbf {\bibinfo {volume} {111}},\ \bibinfo {pages}
  {127003} (\bibinfo {year} {2013})}\BibitemShut {NoStop}%
\bibitem [{\citenamefont {Tabuchi}\ \emph {et~al.}(2014)\citenamefont
  {Tabuchi}, \citenamefont {Ishino}, \citenamefont {Ishikawa}, \citenamefont
  {Yamazaki}, \citenamefont {Usami},\ and\ \citenamefont
  {Nakamura}}]{Tabuchi2014}%
  \BibitemOpen
  \bibfield  {author} {\bibinfo {author} {\bibfnamefont {Y.}~\bibnamefont
  {Tabuchi}}, \bibinfo {author} {\bibfnamefont {S.}~\bibnamefont {Ishino}},
  \bibinfo {author} {\bibfnamefont {T.}~\bibnamefont {Ishikawa}}, \bibinfo
  {author} {\bibfnamefont {R.}~\bibnamefont {Yamazaki}}, \bibinfo {author}
  {\bibfnamefont {K.}~\bibnamefont {Usami}}, \ and\ \bibinfo {author}
  {\bibfnamefont {Y.}~\bibnamefont {Nakamura}},\ }\bibfield  {title} {\enquote
  {\bibinfo {title} {Hybridizing ferromagnetic magnons and microwave photons in
  the quantum limit},}\ }\href {\doibase 10.1103/PhysRevLett.113.083603}
  {\bibfield  {journal} {\bibinfo  {journal} {Phys. Rev. Lett.}\ }\textbf
  {\bibinfo {volume} {113}},\ \bibinfo {pages} {083603} (\bibinfo {year}
  {2014})}\BibitemShut {NoStop}%
\bibitem [{\citenamefont {Zhang}\ \emph {et~al.}(2014)\citenamefont {Zhang},
  \citenamefont {Zou}, \citenamefont {Jiang},\ and\ \citenamefont
  {Tang}}]{Zhang2014}%
  \BibitemOpen
  \bibfield  {author} {\bibinfo {author} {\bibfnamefont {X.}~\bibnamefont
  {Zhang}}, \bibinfo {author} {\bibfnamefont {C.-L.}\ \bibnamefont {Zou}},
  \bibinfo {author} {\bibfnamefont {L.}~\bibnamefont {Jiang}}, \ and\ \bibinfo
  {author} {\bibfnamefont {H.~X.}\ \bibnamefont {Tang}},\ }\bibfield  {title}
  {\enquote {\bibinfo {title} {Strongly coupled magnons and cavity microwave
  photons},}\ }\href {\doibase 10.1103/PhysRevLett.113.156401} {\bibfield
  {journal} {\bibinfo  {journal} {Phys. Rev. Lett.}\ }\textbf {\bibinfo
  {volume} {113}},\ \bibinfo {pages} {156401} (\bibinfo {year}
  {2014})}\BibitemShut {NoStop}%
\bibitem [{\citenamefont {Zhang}\ \emph {et~al.}(2015)\citenamefont {Zhang},
  \citenamefont {Wang}, \citenamefont {Li}, \citenamefont {Luo}, \citenamefont
  {Wu}, \citenamefont {Nori},\ and\ \citenamefont {You}}]{Zhang2015}%
  \BibitemOpen
  \bibfield  {author} {\bibinfo {author} {\bibfnamefont {D.}~\bibnamefont
  {Zhang}}, \bibinfo {author} {\bibfnamefont {X.-M.}\ \bibnamefont {Wang}},
  \bibinfo {author} {\bibfnamefont {T.-F.}\ \bibnamefont {Li}}, \bibinfo
  {author} {\bibfnamefont {X.-Q.}\ \bibnamefont {Luo}}, \bibinfo {author}
  {\bibfnamefont {W.}~\bibnamefont {Wu}}, \bibinfo {author} {\bibfnamefont
  {F.}~\bibnamefont {Nori}}, \ and\ \bibinfo {author} {\bibfnamefont
  {J.}~\bibnamefont {You}},\ }\bibfield  {title} {\enquote {\bibinfo {title}
  {Cavity quantum electrodynamics with ferromagnetic magnons in a small
  yttrium-iron-garnet sphere},}\ }\href
  {http://dx.doi.org/10.1038/npjqi.2015.14} {\bibfield  {journal} {\bibinfo
  {journal} {npj Quantum Information}\ }\textbf {\bibinfo {volume} {1}},\
  \bibinfo {pages} {15014} (\bibinfo {year} {2015})}\BibitemShut {NoStop}%
\bibitem [{\citenamefont {Morris}\ \emph {et~al.}(2017)\citenamefont {Morris},
  \citenamefont {van Loo}, \citenamefont {Kosen},\ and\ \citenamefont
  {Karenowska}}]{Morris2017}%
  \BibitemOpen
  \bibfield  {author} {\bibinfo {author} {\bibfnamefont {R.~G.~E.}\
  \bibnamefont {Morris}}, \bibinfo {author} {\bibfnamefont {A.~F.}\
  \bibnamefont {van Loo}}, \bibinfo {author} {\bibfnamefont {S.}~\bibnamefont
  {Kosen}}, \ and\ \bibinfo {author} {\bibfnamefont {A.~D.}\ \bibnamefont
  {Karenowska}},\ }\bibfield  {title} {\enquote {\bibinfo {title} {Strong
  coupling of magnons in a {YIG} sphere to photons in a planar superconducting
  resonator in the quantum limit},}\ }\href
  {https://doi.org/10.1038/s41598-017-11835-4} {\bibfield  {journal} {\bibinfo
  {journal} {Scientific Reports}\ }\textbf {\bibinfo {volume} {7}},\ \bibinfo
  {pages} {11511} (\bibinfo {year} {2017})}\BibitemShut {NoStop}%
\bibitem [{\citenamefont {Boventer}\ \emph {et~al.}(2018)\citenamefont
  {Boventer}, \citenamefont {Pfirrmann}, \citenamefont {Krause}, \citenamefont
  {Sch\"on}, \citenamefont {Kl\"aui},\ and\ \citenamefont
  {Weides}}]{Boventer2018}%
  \BibitemOpen
  \bibfield  {author} {\bibinfo {author} {\bibfnamefont {I.}~\bibnamefont
  {Boventer}}, \bibinfo {author} {\bibfnamefont {M.}~\bibnamefont {Pfirrmann}},
  \bibinfo {author} {\bibfnamefont {J.}~\bibnamefont {Krause}}, \bibinfo
  {author} {\bibfnamefont {Y.}~\bibnamefont {Sch\"on}}, \bibinfo {author}
  {\bibfnamefont {M.}~\bibnamefont {Kl\"aui}}, \ and\ \bibinfo {author}
  {\bibfnamefont {M.}~\bibnamefont {Weides}},\ }\bibfield  {title} {\enquote
  {\bibinfo {title} {Complex temperature dependence of coupling and dissipation
  of cavity magnon polaritons from millikelvin to room temperature},}\ }\href
  {\doibase 10.1103/PhysRevB.97.184420} {\bibfield  {journal} {\bibinfo
  {journal} {Phys. Rev. B}\ }\textbf {\bibinfo {volume} {97}},\ \bibinfo
  {pages} {184420} (\bibinfo {year} {2018})}\BibitemShut {NoStop}%
\bibitem [{\citenamefont {Boventer}\ \emph {et~al.}(2020)\citenamefont
  {Boventer}, \citenamefont {D\"orflinger}, \citenamefont {Wolz}, \citenamefont
  {Mac\^edo}, \citenamefont {Lebrun}, \citenamefont {Kl\"aui},\ and\
  \citenamefont {Weides}}]{Boventer2020}%
  \BibitemOpen
  \bibfield  {author} {\bibinfo {author} {\bibfnamefont {I.}~\bibnamefont
  {Boventer}}, \bibinfo {author} {\bibfnamefont {C.}~\bibnamefont
  {D\"orflinger}}, \bibinfo {author} {\bibfnamefont {T.}~\bibnamefont {Wolz}},
  \bibinfo {author} {\bibfnamefont {R.}~\bibnamefont {Mac\^edo}}, \bibinfo
  {author} {\bibfnamefont {R.}~\bibnamefont {Lebrun}}, \bibinfo {author}
  {\bibfnamefont {M.}~\bibnamefont {Kl\"aui}}, \ and\ \bibinfo {author}
  {\bibfnamefont {M.}~\bibnamefont {Weides}},\ }\bibfield  {title} {\enquote
  {\bibinfo {title} {Control of the coupling strength and linewidth of a cavity
  magnon-polariton},}\ }\href {\doibase 10.1103/PhysRevResearch.2.013154}
  {\bibfield  {journal} {\bibinfo  {journal} {Phys. Rev. Research}\ }\textbf
  {\bibinfo {volume} {2}},\ \bibinfo {pages} {013154} (\bibinfo {year}
  {2020})}\BibitemShut {NoStop}%
\bibitem [{\citenamefont {Trempler}\ \emph {et~al.}(2020)\citenamefont
  {Trempler}, \citenamefont {Dreyer}, \citenamefont {Geyer}, \citenamefont
  {Hauser}, \citenamefont {Woltersdorf},\ and\ \citenamefont
  {Schmidt}}]{Trempler2020}%
  \BibitemOpen
  \bibfield  {author} {\bibinfo {author} {\bibfnamefont {P.}~\bibnamefont
  {Trempler}}, \bibinfo {author} {\bibfnamefont {R.}~\bibnamefont {Dreyer}},
  \bibinfo {author} {\bibfnamefont {P.}~\bibnamefont {Geyer}}, \bibinfo
  {author} {\bibfnamefont {C.}~\bibnamefont {Hauser}}, \bibinfo {author}
  {\bibfnamefont {G.}~\bibnamefont {Woltersdorf}}, \ and\ \bibinfo {author}
  {\bibfnamefont {G.}~\bibnamefont {Schmidt}},\ }\bibfield  {title} {\enquote
  {\bibinfo {title} {Integration and characterization of micron-sized {YIG}
  structures with very low {Gilbert} damping on arbitrary substrates},}\ }\href
  {\doibase 10.1063/5.0026120} {\bibfield  {journal} {\bibinfo  {journal}
  {Applied Physics Letters}\ }\textbf {\bibinfo {volume} {117}},\ \bibinfo
  {pages} {232401} (\bibinfo {year} {2020})}\BibitemShut {NoStop}%
\bibitem [{\citenamefont {Tabuchi}\ \emph {et~al.}(2015)\citenamefont
  {Tabuchi}, \citenamefont {Ishino}, \citenamefont {Noguchi}, \citenamefont
  {Ishikawa}, \citenamefont {Yamazaki}, \citenamefont {Usami},\ and\
  \citenamefont {Nakamura}}]{Tabuchi2015}%
  \BibitemOpen
  \bibfield  {author} {\bibinfo {author} {\bibfnamefont {Y.}~\bibnamefont
  {Tabuchi}}, \bibinfo {author} {\bibfnamefont {S.}~\bibnamefont {Ishino}},
  \bibinfo {author} {\bibfnamefont {A.}~\bibnamefont {Noguchi}}, \bibinfo
  {author} {\bibfnamefont {T.}~\bibnamefont {Ishikawa}}, \bibinfo {author}
  {\bibfnamefont {R.}~\bibnamefont {Yamazaki}}, \bibinfo {author}
  {\bibfnamefont {K.}~\bibnamefont {Usami}}, \ and\ \bibinfo {author}
  {\bibfnamefont {Y.}~\bibnamefont {Nakamura}},\ }\bibfield  {title} {\enquote
  {\bibinfo {title} {Coherent coupling between a ferromagnetic magnon and a
  superconducting qubit},}\ }\href {\doibase 10.1126/science.aaa3693}
  {\bibfield  {journal} {\bibinfo  {journal} {Science}\ }\textbf {\bibinfo
  {volume} {349}},\ \bibinfo {pages} {405--408} (\bibinfo {year}
  {2015})}\BibitemShut {NoStop}%
\bibitem [{\citenamefont {Li}\ \emph {et~al.}(2019)\citenamefont {Li},
  \citenamefont {Polakovic}, \citenamefont {Wang}, \citenamefont {Xu},
  \citenamefont {Lendinez}, \citenamefont {Zhang}, \citenamefont {Ding},
  \citenamefont {Khaire}, \citenamefont {Saglam}, \citenamefont {Divan},
  \citenamefont {Pearson}, \citenamefont {Kwok}, \citenamefont {Xiao},
  \citenamefont {Novosad}, \citenamefont {Hoffmann},\ and\ \citenamefont
  {Zhang}}]{Li2019}%
  \BibitemOpen
  \bibfield  {author} {\bibinfo {author} {\bibfnamefont {Y.}~\bibnamefont
  {Li}}, \bibinfo {author} {\bibfnamefont {T.}~\bibnamefont {Polakovic}},
  \bibinfo {author} {\bibfnamefont {Y.-L.}\ \bibnamefont {Wang}}, \bibinfo
  {author} {\bibfnamefont {J.}~\bibnamefont {Xu}}, \bibinfo {author}
  {\bibfnamefont {S.}~\bibnamefont {Lendinez}}, \bibinfo {author}
  {\bibfnamefont {Z.}~\bibnamefont {Zhang}}, \bibinfo {author} {\bibfnamefont
  {J.}~\bibnamefont {Ding}}, \bibinfo {author} {\bibfnamefont {T.}~\bibnamefont
  {Khaire}}, \bibinfo {author} {\bibfnamefont {H.}~\bibnamefont {Saglam}},
  \bibinfo {author} {\bibfnamefont {R.}~\bibnamefont {Divan}}, \bibinfo
  {author} {\bibfnamefont {J.}~\bibnamefont {Pearson}}, \bibinfo {author}
  {\bibfnamefont {W.-K.}\ \bibnamefont {Kwok}}, \bibinfo {author}
  {\bibfnamefont {Z.}~\bibnamefont {Xiao}}, \bibinfo {author} {\bibfnamefont
  {V.}~\bibnamefont {Novosad}}, \bibinfo {author} {\bibfnamefont
  {A.}~\bibnamefont {Hoffmann}}, \ and\ \bibinfo {author} {\bibfnamefont
  {W.}~\bibnamefont {Zhang}},\ }\bibfield  {title} {\enquote {\bibinfo {title}
  {Strong coupling between magnons and microwave photons in on-chip
  ferromagnet-superconductor thin-film devices},}\ }\href {\doibase
  10.1103/PhysRevLett.123.107701} {\bibfield  {journal} {\bibinfo  {journal}
  {Phys. Rev. Lett.}\ }\textbf {\bibinfo {volume} {123}},\ \bibinfo {pages}
  {107701} (\bibinfo {year} {2019})}\BibitemShut {NoStop}%
\bibitem [{\citenamefont {Cherepanov}, \citenamefont {Kolokolov},\ and\
  \citenamefont {L'vov}(1993)}]{Cherepanov1993}%
  \BibitemOpen
  \bibfield  {author} {\bibinfo {author} {\bibfnamefont {V.}~\bibnamefont
  {Cherepanov}}, \bibinfo {author} {\bibfnamefont {I.}~\bibnamefont
  {Kolokolov}}, \ and\ \bibinfo {author} {\bibfnamefont {V.}~\bibnamefont
  {L'vov}},\ }\bibfield  {title} {\enquote {\bibinfo {title} {{The saga of YIG:
  Spectra, thermodynamics, interaction and relaxation of magnons in a complex
  magnet}},}\ }\href
  {http://linkinghub.elsevier.com/retrieve/pii/037015739390107O} {\bibfield
  {journal} {\bibinfo  {journal} {Physics Reports}\ }\textbf {\bibinfo {volume}
  {229}},\ \bibinfo {pages} {81--144} (\bibinfo {year} {1993})}\BibitemShut
  {NoStop}%
\bibitem [{\citenamefont {Kosen}\ \emph {et~al.}(2019)\citenamefont {Kosen},
  \citenamefont {van Loo}, \citenamefont {Bozhko}, \citenamefont {Mihalceanu},\
  and\ \citenamefont {Karenowska}}]{Kosen2019}%
  \BibitemOpen
  \bibfield  {author} {\bibinfo {author} {\bibfnamefont {S.}~\bibnamefont
  {Kosen}}, \bibinfo {author} {\bibfnamefont {A.~F.}\ \bibnamefont {van Loo}},
  \bibinfo {author} {\bibfnamefont {D.~A.}\ \bibnamefont {Bozhko}}, \bibinfo
  {author} {\bibfnamefont {L.}~\bibnamefont {Mihalceanu}}, \ and\ \bibinfo
  {author} {\bibfnamefont {A.~D.}\ \bibnamefont {Karenowska}},\ }\bibfield
  {title} {\enquote {\bibinfo {title} {Microwave magnon damping in {YIG} films
  at millikelvin temperatures},}\ }\href {\doibase 10.1063/1.5115266}
  {\bibfield  {journal} {\bibinfo  {journal} {APL Materials}\ }\textbf
  {\bibinfo {volume} {7}},\ \bibinfo {pages} {101120} (\bibinfo {year}
  {2019})}\BibitemShut {NoStop}%
\bibitem [{\citenamefont {Mihalceanu}\ \emph {et~al.}(2018)\citenamefont
  {Mihalceanu}, \citenamefont {Vasyuchka}, \citenamefont {Bozhko},
  \citenamefont {Langner}, \citenamefont {Nechiporuk}, \citenamefont
  {Romanyuk}, \citenamefont {Hillebrands},\ and\ \citenamefont
  {Serga}}]{Mihalceanu2018}%
  \BibitemOpen
  \bibfield  {author} {\bibinfo {author} {\bibfnamefont {L.}~\bibnamefont
  {Mihalceanu}}, \bibinfo {author} {\bibfnamefont {V.~I.}\ \bibnamefont
  {Vasyuchka}}, \bibinfo {author} {\bibfnamefont {D.~A.}\ \bibnamefont
  {Bozhko}}, \bibinfo {author} {\bibfnamefont {T.}~\bibnamefont {Langner}},
  \bibinfo {author} {\bibfnamefont {A.~Y.}\ \bibnamefont {Nechiporuk}},
  \bibinfo {author} {\bibfnamefont {V.~F.}\ \bibnamefont {Romanyuk}}, \bibinfo
  {author} {\bibfnamefont {B.}~\bibnamefont {Hillebrands}}, \ and\ \bibinfo
  {author} {\bibfnamefont {A.~A.}\ \bibnamefont {Serga}},\ }\bibfield  {title}
  {\enquote {\bibinfo {title} {{Temperature-dependent relaxation of
  dipole-exchange magnons in yttrium iron garnet films}},}\ }\href {\doibase
  10.1103/PhysRevB.97.214405} {\bibfield  {journal} {\bibinfo  {journal} {Phys.
  Rev. B}\ }\textbf {\bibinfo {volume} {97}},\ \bibinfo {pages} {214405}
  (\bibinfo {year} {2018})}\BibitemShut {NoStop}%
\bibitem [{\citenamefont {Kruglyak}, \citenamefont {Demokritov},\ and\
  \citenamefont {Grundler}(2010)}]{Kruglyak2010}%
  \BibitemOpen
  \bibfield  {author} {\bibinfo {author} {\bibfnamefont {V.~V.}\ \bibnamefont
  {Kruglyak}}, \bibinfo {author} {\bibfnamefont {S.~O.}\ \bibnamefont
  {Demokritov}}, \ and\ \bibinfo {author} {\bibfnamefont {D.}~\bibnamefont
  {Grundler}},\ }\bibfield  {title} {\enquote {\bibinfo {title} {Magnonics},}\
  }\href {http://stacks.iop.org/0022-3727/43/i=26/a=264001} {\bibfield
  {journal} {\bibinfo  {journal} {Journal of Physics D: Applied Physics}\
  }\textbf {\bibinfo {volume} {43}},\ \bibinfo {pages} {264001} (\bibinfo
  {year} {2010})}\BibitemShut {NoStop}%
\bibitem [{\citenamefont {Serga}, \citenamefont {Chumak},\ and\ \citenamefont
  {Hillebrands}(2010)}]{Serga2010}%
  \BibitemOpen
  \bibfield  {author} {\bibinfo {author} {\bibfnamefont {A.~A.}\ \bibnamefont
  {Serga}}, \bibinfo {author} {\bibfnamefont {A.~V.}\ \bibnamefont {Chumak}}, \
  and\ \bibinfo {author} {\bibfnamefont {B.}~\bibnamefont {Hillebrands}},\
  }\bibfield  {title} {\enquote {\bibinfo {title} {{YIG} magnonics},}\ }\href
  {\doibase 10.1088/0022-3727/43/26/264002} {\bibfield  {journal} {\bibinfo
  {journal} {Journal of Physics D: Applied Physics}\ }\textbf {\bibinfo
  {volume} {43}},\ \bibinfo {pages} {264002} (\bibinfo {year}
  {2010})}\BibitemShut {NoStop}%
\bibitem [{\citenamefont {Lenk}\ \emph {et~al.}(2011)\citenamefont {Lenk},
  \citenamefont {Ulrichs}, \citenamefont {Garbs},\ and\ \citenamefont
  {Münzenberg}}]{Lenk2011}%
  \BibitemOpen
  \bibfield  {author} {\bibinfo {author} {\bibfnamefont {B.}~\bibnamefont
  {Lenk}}, \bibinfo {author} {\bibfnamefont {H.}~\bibnamefont {Ulrichs}},
  \bibinfo {author} {\bibfnamefont {F.}~\bibnamefont {Garbs}}, \ and\ \bibinfo
  {author} {\bibfnamefont {M.}~\bibnamefont {Münzenberg}},\ }\bibfield
  {title} {\enquote {\bibinfo {title} {The building blocks of magnonics},}\
  }\href {\doibase https://doi.org/10.1016/j.physrep.2011.06.003} {\bibfield
  {journal} {\bibinfo  {journal} {Physics Reports}\ }\textbf {\bibinfo {volume}
  {507}},\ \bibinfo {pages} {107 -- 136} (\bibinfo {year} {2011})}\BibitemShut
  {NoStop}%
\bibitem [{\citenamefont {Fischer}\ \emph {et~al.}(2017)\citenamefont
  {Fischer}, \citenamefont {Kewenig}, \citenamefont {Bozhko}, \citenamefont
  {Serga}, \citenamefont {Syvorotka}, \citenamefont {Ciubotaru}, \citenamefont
  {Adelmann}, \citenamefont {Hillebrands},\ and\ \citenamefont
  {Chumak}}]{Fischer2017}%
  \BibitemOpen
  \bibfield  {author} {\bibinfo {author} {\bibfnamefont {T.}~\bibnamefont
  {Fischer}}, \bibinfo {author} {\bibfnamefont {M.}~\bibnamefont {Kewenig}},
  \bibinfo {author} {\bibfnamefont {D.~A.}\ \bibnamefont {Bozhko}}, \bibinfo
  {author} {\bibfnamefont {A.~A.}\ \bibnamefont {Serga}}, \bibinfo {author}
  {\bibfnamefont {I.~I.}\ \bibnamefont {Syvorotka}}, \bibinfo {author}
  {\bibfnamefont {F.}~\bibnamefont {Ciubotaru}}, \bibinfo {author}
  {\bibfnamefont {C.}~\bibnamefont {Adelmann}}, \bibinfo {author}
  {\bibfnamefont {B.}~\bibnamefont {Hillebrands}}, \ and\ \bibinfo {author}
  {\bibfnamefont {A.~V.}\ \bibnamefont {Chumak}},\ }\bibfield  {title}
  {\enquote {\bibinfo {title} {{Experimental prototype of a spin-wave majority
  gate}},}\ }\href {\doibase 10.1063/1.4979840} {\bibfield  {journal} {\bibinfo
   {journal} {Applied Physics Letters}\ }\textbf {\bibinfo {volume} {110}},\
  \bibinfo {pages} {152401} (\bibinfo {year} {2017})}\BibitemShut {NoStop}%
\bibitem [{\citenamefont {Frey}\ \emph {et~al.}(2020)\citenamefont {Frey},
  \citenamefont {Nikitin}, \citenamefont {Bozhko}, \citenamefont {Bunyaev},
  \citenamefont {Kakazei}, \citenamefont {Ustinov}, \citenamefont {Kalinikos},
  \citenamefont {Ciubotaru}, \citenamefont {Chumak}, \citenamefont {Wang},
  \citenamefont {Tiberkevich}, \citenamefont {Hillebrands},\ and\ \citenamefont
  {Serga}}]{Frey2020}%
  \BibitemOpen
  \bibfield  {author} {\bibinfo {author} {\bibfnamefont {P.}~\bibnamefont
  {Frey}}, \bibinfo {author} {\bibfnamefont {A.~A.}\ \bibnamefont {Nikitin}},
  \bibinfo {author} {\bibfnamefont {D.~A.}\ \bibnamefont {Bozhko}}, \bibinfo
  {author} {\bibfnamefont {S.~A.}\ \bibnamefont {Bunyaev}}, \bibinfo {author}
  {\bibfnamefont {G.~N.}\ \bibnamefont {Kakazei}}, \bibinfo {author}
  {\bibfnamefont {A.~B.}\ \bibnamefont {Ustinov}}, \bibinfo {author}
  {\bibfnamefont {B.~A.}\ \bibnamefont {Kalinikos}}, \bibinfo {author}
  {\bibfnamefont {F.}~\bibnamefont {Ciubotaru}}, \bibinfo {author}
  {\bibfnamefont {A.~V.}\ \bibnamefont {Chumak}}, \bibinfo {author}
  {\bibfnamefont {Q.}~\bibnamefont {Wang}}, \bibinfo {author} {\bibfnamefont
  {V.~S.}\ \bibnamefont {Tiberkevich}}, \bibinfo {author} {\bibfnamefont
  {B.}~\bibnamefont {Hillebrands}}, \ and\ \bibinfo {author} {\bibfnamefont
  {A.~A.}\ \bibnamefont {Serga}},\ }\bibfield  {title} {\enquote {\bibinfo
  {title} {{Reflection-less width-modulated magnonic crystal}},}\ }\href
  {\doibase 10.1038/s42005-020-0281-y} {\bibfield  {journal} {\bibinfo
  {journal} {Communications Physics}\ }\textbf {\bibinfo {volume} {3}},\
  \bibinfo {pages} {17} (\bibinfo {year} {2020})}\BibitemShut {NoStop}%
\bibitem [{\citenamefont {Prokopenko}\ \emph {et~al.}(2019)\citenamefont
  {Prokopenko}, \citenamefont {Bozhko}, \citenamefont {Tyberkevych},
  \citenamefont {Chumak}, \citenamefont {Vasyuchka}, \citenamefont {Serga},
  \citenamefont {Dzyapko}, \citenamefont {Verba}, \citenamefont {Talalaevskij},
  \citenamefont {Slobodianiuk}, \citenamefont {Kobljanskyj}, \citenamefont
  {Moiseienko}, \citenamefont {Sholom},\ and\ \citenamefont
  {Malyshev}}]{Prokopenko2019}%
  \BibitemOpen
  \bibfield  {author} {\bibinfo {author} {\bibfnamefont {O.~V.}\ \bibnamefont
  {Prokopenko}}, \bibinfo {author} {\bibfnamefont {D.~A.}\ \bibnamefont
  {Bozhko}}, \bibinfo {author} {\bibfnamefont {V.~S.}\ \bibnamefont
  {Tyberkevych}}, \bibinfo {author} {\bibfnamefont {A.~V.}\ \bibnamefont
  {Chumak}}, \bibinfo {author} {\bibfnamefont {V.~I.}\ \bibnamefont
  {Vasyuchka}}, \bibinfo {author} {\bibfnamefont {A.~A.}\ \bibnamefont
  {Serga}}, \bibinfo {author} {\bibfnamefont {O.}~\bibnamefont {Dzyapko}},
  \bibinfo {author} {\bibfnamefont {R.~V.}\ \bibnamefont {Verba}}, \bibinfo
  {author} {\bibfnamefont {A.~V.}\ \bibnamefont {Talalaevskij}}, \bibinfo
  {author} {\bibfnamefont {D.~V.}\ \bibnamefont {Slobodianiuk}}, \bibinfo
  {author} {\bibfnamefont {Y.~V.}\ \bibnamefont {Kobljanskyj}}, \bibinfo
  {author} {\bibfnamefont {V.~A.}\ \bibnamefont {Moiseienko}}, \bibinfo
  {author} {\bibfnamefont {S.~V.}\ \bibnamefont {Sholom}}, \ and\ \bibinfo
  {author} {\bibfnamefont {V.~Y.}\ \bibnamefont {Malyshev}},\ }\bibfield
  {title} {\enquote {\bibinfo {title} {{Recent Trends in Microwave Magnetism
  and Superconductivity}},}\ }\href {\doibase 10.15407/ujpe64.10.888}
  {\bibfield  {journal} {\bibinfo  {journal} {Ukrainian Journal of Physics}\
  }\textbf {\bibinfo {volume} {64}},\ \bibinfo {pages} {888} (\bibinfo {year}
  {2019})}\BibitemShut {NoStop}%
\bibitem [{\citenamefont {Aharoni}(1998)}]{Aharoni1998}%
  \BibitemOpen
  \bibfield  {author} {\bibinfo {author} {\bibfnamefont {A.}~\bibnamefont
  {Aharoni}},\ }\bibfield  {title} {\enquote {\bibinfo {title} {Demagnetizing
  factors for rectangular ferromagnetic prisms},}\ }\href {\doibase
  10.1063/1.367113} {\bibfield  {journal} {\bibinfo  {journal} {Journal of
  Applied Physics}\ }\textbf {\bibinfo {volume} {83}},\ \bibinfo {pages}
  {3432--3434} (\bibinfo {year} {1998})}\BibitemShut {NoStop}%
\bibitem [{\citenamefont {Banerjee}\ \emph {et~al.}(2018)\citenamefont
  {Banerjee}, \citenamefont {Heath}, \citenamefont {Morozov}, \citenamefont
  {Hemakumara}, \citenamefont {Nasti}, \citenamefont {Thayne},\ and\
  \citenamefont {Hadfield}}]{Banerjee2018}%
  \BibitemOpen
  \bibfield  {author} {\bibinfo {author} {\bibfnamefont {A.}~\bibnamefont
  {Banerjee}}, \bibinfo {author} {\bibfnamefont {R.~M.}\ \bibnamefont {Heath}},
  \bibinfo {author} {\bibfnamefont {D.}~\bibnamefont {Morozov}}, \bibinfo
  {author} {\bibfnamefont {D.}~\bibnamefont {Hemakumara}}, \bibinfo {author}
  {\bibfnamefont {U.}~\bibnamefont {Nasti}}, \bibinfo {author} {\bibfnamefont
  {I.}~\bibnamefont {Thayne}}, \ and\ \bibinfo {author} {\bibfnamefont {R.~H.}\
  \bibnamefont {Hadfield}},\ }\bibfield  {title} {\enquote {\bibinfo {title}
  {Optical properties of refractory metal based thin films},}\ }\href {\doibase
  10.1364/OME.8.002072} {\bibfield  {journal} {\bibinfo  {journal} {Opt. Mater.
  Express}\ }\textbf {\bibinfo {volume} {8}},\ \bibinfo {pages} {2072--2088}
  (\bibinfo {year} {2018})}\BibitemShut {NoStop}%
\bibitem [{\citenamefont {Voesch}\ \emph {et~al.}(2015)\citenamefont {Voesch},
  \citenamefont {Thiemann}, \citenamefont {Bothner}, \citenamefont {Dressel},\
  and\ \citenamefont {Scheffler}}]{Voesch2015}%
  \BibitemOpen
  \bibfield  {author} {\bibinfo {author} {\bibfnamefont {W.}~\bibnamefont
  {Voesch}}, \bibinfo {author} {\bibfnamefont {M.}~\bibnamefont {Thiemann}},
  \bibinfo {author} {\bibfnamefont {D.}~\bibnamefont {Bothner}}, \bibinfo
  {author} {\bibfnamefont {M.}~\bibnamefont {Dressel}}, \ and\ \bibinfo
  {author} {\bibfnamefont {M.}~\bibnamefont {Scheffler}},\ }\bibfield  {title}
  {\enquote {\bibinfo {title} {On-chip {ESR} measurements of {DPPH} at {mK}
  temperatures},}\ }\href {\doibase
  https://doi.org/10.1016/j.phpro.2015.12.063} {\bibfield  {journal} {\bibinfo
  {journal} {Physics Procedia}\ }\textbf {\bibinfo {volume} {75}},\ \bibinfo
  {pages} {503--510} (\bibinfo {year} {2015})},\ \bibinfo {note} {20th
  International Conference on Magnetism, ICM 2015}\BibitemShut {NoStop}%
\bibitem [{\citenamefont {Golovchanskiy}\ \emph {et~al.}(2019)\citenamefont
  {Golovchanskiy}, \citenamefont {Abramov}, \citenamefont {Pfirrmann},
  \citenamefont {Piskor}, \citenamefont {Voss}, \citenamefont {Baranov},
  \citenamefont {Hovhannisyan}, \citenamefont {Stolyarov}, \citenamefont
  {Dubs}, \citenamefont {Golubov}, \citenamefont {Ryazanov}, \citenamefont
  {Ustinov},\ and\ \citenamefont {Weides}}]{Golovchanskiy2019}%
  \BibitemOpen
  \bibfield  {author} {\bibinfo {author} {\bibfnamefont {I.}~\bibnamefont
  {Golovchanskiy}}, \bibinfo {author} {\bibfnamefont {N.}~\bibnamefont
  {Abramov}}, \bibinfo {author} {\bibfnamefont {M.}~\bibnamefont {Pfirrmann}},
  \bibinfo {author} {\bibfnamefont {T.}~\bibnamefont {Piskor}}, \bibinfo
  {author} {\bibfnamefont {J.}~\bibnamefont {Voss}}, \bibinfo {author}
  {\bibfnamefont {D.}~\bibnamefont {Baranov}}, \bibinfo {author} {\bibfnamefont
  {R.}~\bibnamefont {Hovhannisyan}}, \bibinfo {author} {\bibfnamefont
  {V.}~\bibnamefont {Stolyarov}}, \bibinfo {author} {\bibfnamefont
  {C.}~\bibnamefont {Dubs}}, \bibinfo {author} {\bibfnamefont {A.}~\bibnamefont
  {Golubov}}, \bibinfo {author} {\bibfnamefont {V.}~\bibnamefont {Ryazanov}},
  \bibinfo {author} {\bibfnamefont {A.}~\bibnamefont {Ustinov}}, \ and\
  \bibinfo {author} {\bibfnamefont {M.}~\bibnamefont {Weides}},\ }\bibfield
  {title} {\enquote {\bibinfo {title} {Interplay of magnetization dynamics with
  a microwave waveguide at cryogenic temperatures},}\ }\href {\doibase
  10.1103/PhysRevApplied.11.044076} {\bibfield  {journal} {\bibinfo  {journal}
  {Phys. Rev. Applied}\ }\textbf {\bibinfo {volume} {11}},\ \bibinfo {pages}
  {044076} (\bibinfo {year} {2019})}\BibitemShut {NoStop}%
\bibitem [{\citenamefont {Kittel}(1948)}]{Kittel1948}%
  \BibitemOpen
  \bibfield  {author} {\bibinfo {author} {\bibfnamefont {C.}~\bibnamefont
  {Kittel}},\ }\bibfield  {title} {\enquote {\bibinfo {title} {On the theory of
  ferromagnetic resonance absorption},}\ }\href {\doibase
  10.1103/PhysRev.73.155} {\bibfield  {journal} {\bibinfo  {journal} {Phys.
  Rev.}\ }\textbf {\bibinfo {volume} {73}},\ \bibinfo {pages} {155--161}
  (\bibinfo {year} {1948})}\BibitemShut {NoStop}%
\bibitem [{\citenamefont {Fano}(1961)}]{Fano1961}%
  \BibitemOpen
  \bibfield  {author} {\bibinfo {author} {\bibfnamefont {U.}~\bibnamefont
  {Fano}},\ }\bibfield  {title} {\enquote {\bibinfo {title} {Effects of
  configuration interaction on intensities and phase shifts},}\ }\href
  {\doibase 10.1103/PhysRev.124.1866} {\bibfield  {journal} {\bibinfo
  {journal} {Phys. Rev.}\ }\textbf {\bibinfo {volume} {124}},\ \bibinfo {pages}
  {1866--1878} (\bibinfo {year} {1961})}\BibitemShut {NoStop}%
\bibitem [{\citenamefont {Limonov}\ \emph {et~al.}(2017)\citenamefont
  {Limonov}, \citenamefont {Rybin}, \citenamefont {Poddubny},\ and\
  \citenamefont {Kivshar}}]{Limonov2017}%
  \BibitemOpen
  \bibfield  {author} {\bibinfo {author} {\bibfnamefont {M.~F.}\ \bibnamefont
  {Limonov}}, \bibinfo {author} {\bibfnamefont {M.~V.}\ \bibnamefont {Rybin}},
  \bibinfo {author} {\bibfnamefont {A.~N.}\ \bibnamefont {Poddubny}}, \ and\
  \bibinfo {author} {\bibfnamefont {Y.~S.}\ \bibnamefont {Kivshar}},\
  }\bibfield  {title} {\enquote {\bibinfo {title} {Fano resonances in
  photonics},}\ }\href {\doibase 10.1038/nphoton.2017.142} {\bibfield
  {journal} {\bibinfo  {journal} {Nature Photonics}\ }\textbf {\bibinfo
  {volume} {11}},\ \bibinfo {pages} {543--554} (\bibinfo {year}
  {2017})}\BibitemShut {NoStop}%
\bibitem [{\citenamefont {Macêdo}\ \emph {et~al.}(2021)\citenamefont
  {Macêdo}, \citenamefont {Holland}, \citenamefont {Baity}, \citenamefont
  {McLellan}, \citenamefont {Livesey}, \citenamefont {Stamps}, \citenamefont
  {Weides},\ and\ \citenamefont {Bozhko}}]{RairsPaper}%
  \BibitemOpen
  \bibfield  {author} {\bibinfo {author} {\bibfnamefont {R.}~\bibnamefont
  {Macêdo}}, \bibinfo {author} {\bibfnamefont {R.~C.}\ \bibnamefont
  {Holland}}, \bibinfo {author} {\bibfnamefont {P.~G.}\ \bibnamefont {Baity}},
  \bibinfo {author} {\bibfnamefont {L.~J.}\ \bibnamefont {McLellan}}, \bibinfo
  {author} {\bibfnamefont {K.~L.}\ \bibnamefont {Livesey}}, \bibinfo {author}
  {\bibfnamefont {R.~L.}\ \bibnamefont {Stamps}}, \bibinfo {author}
  {\bibfnamefont {M.~P.}\ \bibnamefont {Weides}}, \ and\ \bibinfo {author}
  {\bibfnamefont {D.~A.}\ \bibnamefont {Bozhko}},\ }\bibfield  {title}
  {\enquote {\bibinfo {title} {Electromagnetic approach to cavity
  spintronics},}\ }\href {\doibase 10.1103/PhysRevApplied.15.024065} {\bibfield
   {journal} {\bibinfo  {journal} {Phys. Rev. Applied}\ }\textbf {\bibinfo
  {volume} {15}},\ \bibinfo {pages} {024065} (\bibinfo {year}
  {2021})}\BibitemShut {NoStop}%
\bibitem [{\citenamefont {Bothner}\ \emph {et~al.}(2012)\citenamefont
  {Bothner}, \citenamefont {Gaber}, \citenamefont {Kemmler}, \citenamefont
  {Koelle}, \citenamefont {Kleiner}, \citenamefont {W\"unsch},\ and\
  \citenamefont {Siegel}}]{Bothner2012}%
  \BibitemOpen
  \bibfield  {author} {\bibinfo {author} {\bibfnamefont {D.}~\bibnamefont
  {Bothner}}, \bibinfo {author} {\bibfnamefont {T.}~\bibnamefont {Gaber}},
  \bibinfo {author} {\bibfnamefont {M.}~\bibnamefont {Kemmler}}, \bibinfo
  {author} {\bibfnamefont {D.}~\bibnamefont {Koelle}}, \bibinfo {author}
  {\bibfnamefont {R.}~\bibnamefont {Kleiner}}, \bibinfo {author} {\bibfnamefont
  {S.}~\bibnamefont {W\"unsch}}, \ and\ \bibinfo {author} {\bibfnamefont
  {M.}~\bibnamefont {Siegel}},\ }\bibfield  {title} {\enquote {\bibinfo {title}
  {Magnetic hysteresis effects in superconducting coplanar microwave
  resonators},}\ }\href {\doibase 10.1103/PhysRevB.86.014517} {\bibfield
  {journal} {\bibinfo  {journal} {Phys. Rev. B}\ }\textbf {\bibinfo {volume}
  {86}},\ \bibinfo {pages} {014517} (\bibinfo {year} {2012})}\BibitemShut
  {NoStop}%
\bibitem [{\citenamefont {Maier-Flaig}\ \emph {et~al.}(2017)\citenamefont
  {Maier-Flaig}, \citenamefont {Klingler}, \citenamefont {Dubs}, \citenamefont
  {Surzhenko}, \citenamefont {Gross}, \citenamefont {Weiler}, \citenamefont
  {Huebl},\ and\ \citenamefont {Goennenwein}}]{MaierFlaig2017}%
  \BibitemOpen
  \bibfield  {author} {\bibinfo {author} {\bibfnamefont {H.}~\bibnamefont
  {Maier-Flaig}}, \bibinfo {author} {\bibfnamefont {S.}~\bibnamefont
  {Klingler}}, \bibinfo {author} {\bibfnamefont {C.}~\bibnamefont {Dubs}},
  \bibinfo {author} {\bibfnamefont {O.}~\bibnamefont {Surzhenko}}, \bibinfo
  {author} {\bibfnamefont {R.}~\bibnamefont {Gross}}, \bibinfo {author}
  {\bibfnamefont {M.}~\bibnamefont {Weiler}}, \bibinfo {author} {\bibfnamefont
  {H.}~\bibnamefont {Huebl}}, \ and\ \bibinfo {author} {\bibfnamefont
  {S.~T.~B.}\ \bibnamefont {Goennenwein}},\ }\bibfield  {title} {\enquote
  {\bibinfo {title} {Temperature-dependent magnetic damping of yttrium iron
  garnet spheres},}\ }\href {\doibase 10.1103/PhysRevB.95.214423} {\bibfield
  {journal} {\bibinfo  {journal} {Phys. Rev. B}\ }\textbf {\bibinfo {volume}
  {95}},\ \bibinfo {pages} {214423} (\bibinfo {year} {2017})}\BibitemShut
  {NoStop}%
\bibitem [{\citenamefont {Stancil}\ and\ \citenamefont
  {Prabhakar}(2009)}]{SpinWavesBook}%
  \BibitemOpen
  \bibfield  {author} {\bibinfo {author} {\bibfnamefont {D.~D.}\ \bibnamefont
  {Stancil}}\ and\ \bibinfo {author} {\bibfnamefont {A.}~\bibnamefont
  {Prabhakar}},\ }\href@noop {} {\emph {\bibinfo {title} {Spin Waves: Theory
  and Applications}}}\ (\bibinfo  {publisher} {Springer},\ \bibinfo {year}
  {2009})\BibitemShut {NoStop}%
\bibitem [{\citenamefont {Ross}, \citenamefont {Kostylev},\ and\ \citenamefont
  {Stamps}(2011)}]{Ross2011}%
  \BibitemOpen
  \bibfield  {author} {\bibinfo {author} {\bibfnamefont {N.}~\bibnamefont
  {Ross}}, \bibinfo {author} {\bibfnamefont {M.}~\bibnamefont {Kostylev}}, \
  and\ \bibinfo {author} {\bibfnamefont {R.~L.}\ \bibnamefont {Stamps}},\
  }\bibfield  {title} {\enquote {\bibinfo {title} {Effect of disorder studied
  with ferromagnetic resonance for arrays of tangentially magnetized submicron
  permalloy disks fabricated by nanosphere lithography},}\ }\href {\doibase
  10.1063/1.3526307} {\bibfield  {journal} {\bibinfo  {journal} {Journal of
  Applied Physics}\ }\textbf {\bibinfo {volume} {109}},\ \bibinfo {pages}
  {013906} (\bibinfo {year} {2011})}\BibitemShut {NoStop}%
\bibitem [{\citenamefont {Lee}\ \emph {et~al.}(2016)\citenamefont {Lee},
  \citenamefont {Grudichak}, \citenamefont {Sklenar}, \citenamefont {Tsai},
  \citenamefont {Jang}, \citenamefont {Yang}, \citenamefont {Zhang},\ and\
  \citenamefont {Ketterson}}]{Lee2016}%
  \BibitemOpen
  \bibfield  {author} {\bibinfo {author} {\bibfnamefont {S.}~\bibnamefont
  {Lee}}, \bibinfo {author} {\bibfnamefont {S.}~\bibnamefont {Grudichak}},
  \bibinfo {author} {\bibfnamefont {J.}~\bibnamefont {Sklenar}}, \bibinfo
  {author} {\bibfnamefont {C.~C.}\ \bibnamefont {Tsai}}, \bibinfo {author}
  {\bibfnamefont {M.}~\bibnamefont {Jang}}, \bibinfo {author} {\bibfnamefont
  {Q.}~\bibnamefont {Yang}}, \bibinfo {author} {\bibfnamefont {H.}~\bibnamefont
  {Zhang}}, \ and\ \bibinfo {author} {\bibfnamefont {J.~B.}\ \bibnamefont
  {Ketterson}},\ }\bibfield  {title} {\enquote {\bibinfo {title} {Ferromagnetic
  resonance of a {YIG} film in the low frequency regime},}\ }\href {\doibase
  10.1063/1.4956435} {\bibfield  {journal} {\bibinfo  {journal} {Journal of
  Applied Physics}\ }\textbf {\bibinfo {volume} {120}},\ \bibinfo {pages}
  {033905} (\bibinfo {year} {2016})}\BibitemShut {NoStop}%
\bibitem [{\citenamefont {Algra}\ and\ \citenamefont
  {Hansen}(1982)}]{Algra1982}%
  \BibitemOpen
  \bibfield  {author} {\bibinfo {author} {\bibfnamefont {H.~A.}\ \bibnamefont
  {Algra}}\ and\ \bibinfo {author} {\bibfnamefont {P.}~\bibnamefont {Hansen}},\
  }\bibfield  {title} {\enquote {\bibinfo {title} {{Temperature dependence of
  the saturation magnetization of ion-implanted YIG films}},}\ }\href {\doibase
  10.1007/BF00632432} {\bibfield  {journal} {\bibinfo  {journal} {Applied
  Physics A}\ }\textbf {\bibinfo {volume} {29}},\ \bibinfo {pages} {83--86}
  (\bibinfo {year} {1982})}\BibitemShut {NoStop}%
\bibitem [{\citenamefont {Agrawal}\ \emph {et~al.}(2013)\citenamefont
  {Agrawal}, \citenamefont {Vasyuchka}, \citenamefont {Serga}, \citenamefont
  {Karenowska}, \citenamefont {Melkov},\ and\ \citenamefont
  {Hillebrands}}]{Agrawal2013}%
  \BibitemOpen
  \bibfield  {author} {\bibinfo {author} {\bibfnamefont {M.}~\bibnamefont
  {Agrawal}}, \bibinfo {author} {\bibfnamefont {V.~I.}\ \bibnamefont
  {Vasyuchka}}, \bibinfo {author} {\bibfnamefont {A.~A.}\ \bibnamefont
  {Serga}}, \bibinfo {author} {\bibfnamefont {A.~D.}\ \bibnamefont
  {Karenowska}}, \bibinfo {author} {\bibfnamefont {G.~A.}\ \bibnamefont
  {Melkov}}, \ and\ \bibinfo {author} {\bibfnamefont {B.}~\bibnamefont
  {Hillebrands}},\ }\bibfield  {title} {\enquote {\bibinfo {title} {{Direct
  Measurement of Magnon Temperature: New Insight into Magnon-Phonon Coupling in
  Magnetic Insulators}},}\ }\href {\doibase 10.1103/PhysRevLett.111.107204}
  {\bibfield  {journal} {\bibinfo  {journal} {Phys. Rev. Lett.}\ }\textbf
  {\bibinfo {volume} {111}},\ \bibinfo {pages} {107204} (\bibinfo {year}
  {2013})}\BibitemShut {NoStop}%
\bibitem [{\citenamefont {Kostylev}\ \emph {et~al.}(2005)\citenamefont
  {Kostylev}, \citenamefont {Serga}, \citenamefont {Schneider}, \citenamefont
  {Leven},\ and\ \citenamefont {Hillebrands}}]{Kostylev2005}%
  \BibitemOpen
  \bibfield  {author} {\bibinfo {author} {\bibfnamefont {M.~P.}\ \bibnamefont
  {Kostylev}}, \bibinfo {author} {\bibfnamefont {A.~A.}\ \bibnamefont {Serga}},
  \bibinfo {author} {\bibfnamefont {T.}~\bibnamefont {Schneider}}, \bibinfo
  {author} {\bibfnamefont {B.}~\bibnamefont {Leven}}, \ and\ \bibinfo {author}
  {\bibfnamefont {B.}~\bibnamefont {Hillebrands}},\ }\bibfield  {title}
  {\enquote {\bibinfo {title} {Spin-wave logical gates},}\ }\href {\doibase
  10.1063/1.2089147} {\bibfield  {journal} {\bibinfo  {journal} {Applied
  Physics Letters}\ }\textbf {\bibinfo {volume} {87}},\ \bibinfo {pages}
  {153501} (\bibinfo {year} {2005})}\BibitemShut {NoStop}%
\bibitem [{\citenamefont {Vogt}\ \emph {et~al.}(2014)\citenamefont {Vogt},
  \citenamefont {Fradin}, \citenamefont {Pearson}, \citenamefont {Sebastian},
  \citenamefont {Bader}, \citenamefont {Hillebrands}, \citenamefont
  {Hoffmann},\ and\ \citenamefont {Schultheiss}}]{Vogt2014}%
  \BibitemOpen
  \bibfield  {author} {\bibinfo {author} {\bibfnamefont {K.}~\bibnamefont
  {Vogt}}, \bibinfo {author} {\bibfnamefont {F.~Y.}\ \bibnamefont {Fradin}},
  \bibinfo {author} {\bibfnamefont {J.~E.}\ \bibnamefont {Pearson}}, \bibinfo
  {author} {\bibfnamefont {T.}~\bibnamefont {Sebastian}}, \bibinfo {author}
  {\bibfnamefont {S.~D.}\ \bibnamefont {Bader}}, \bibinfo {author}
  {\bibfnamefont {B.}~\bibnamefont {Hillebrands}}, \bibinfo {author}
  {\bibfnamefont {A.}~\bibnamefont {Hoffmann}}, \ and\ \bibinfo {author}
  {\bibfnamefont {H.}~\bibnamefont {Schultheiss}},\ }\bibfield  {title}
  {\enquote {\bibinfo {title} {Realization of a spin-wave multiplexer},}\
  }\href {\doibase 10.1038/ncomms4727} {\bibfield  {journal} {\bibinfo
  {journal} {Nature Communications}\ }\textbf {\bibinfo {volume} {5}},\
  \bibinfo {pages} {3727} (\bibinfo {year} {2014})}\BibitemShut {NoStop}%
\bibitem [{\citenamefont {Green}(1971)}]{Green1971}%
  \BibitemOpen
  \bibfield  {author} {\bibinfo {author} {\bibfnamefont {M.~A.}\ \bibnamefont
  {Green}},\ }\bibfield  {title} {\enquote {\bibinfo {title} {Residual fields
  in superconducting dipole and quadrupole magnets},}\ }\href {\doibase
  10.1109/TNS.1971.4326146} {\bibfield  {journal} {\bibinfo  {journal} {IEEE
  Transactions on Nuclear Science}\ }\textbf {\bibinfo {volume} {18}},\
  \bibinfo {pages} {664--668} (\bibinfo {year} {1971})}\BibitemShut {NoStop}%
\bibitem [{\citenamefont {Schneider}(2020)}]{Schneider2020}%
  \BibitemOpen
  \bibfield  {author} {\bibinfo {author} {\bibfnamefont {A.}~\bibnamefont
  {Schneider}},\ }\emph {\bibinfo {title} {Quantum Sensing Experiments with
  Superconducting Qubits}},\ \href@noop {} {Ph.D. thesis},\ \bibinfo  {school}
  {Karlsruher Instituts für Technologie} (\bibinfo {year} {2020})\BibitemShut
  {NoStop}%
\end{thebibliography}%

\clearpage
\onecolumngrid

\section{Supplementary material for "Strong magnon-photon coupling with chip-integrated YIG in the zero-temperature limit"}

\beginsupplement

\begin{figure*}[p]
\centering
\includegraphics[scale=0.7]{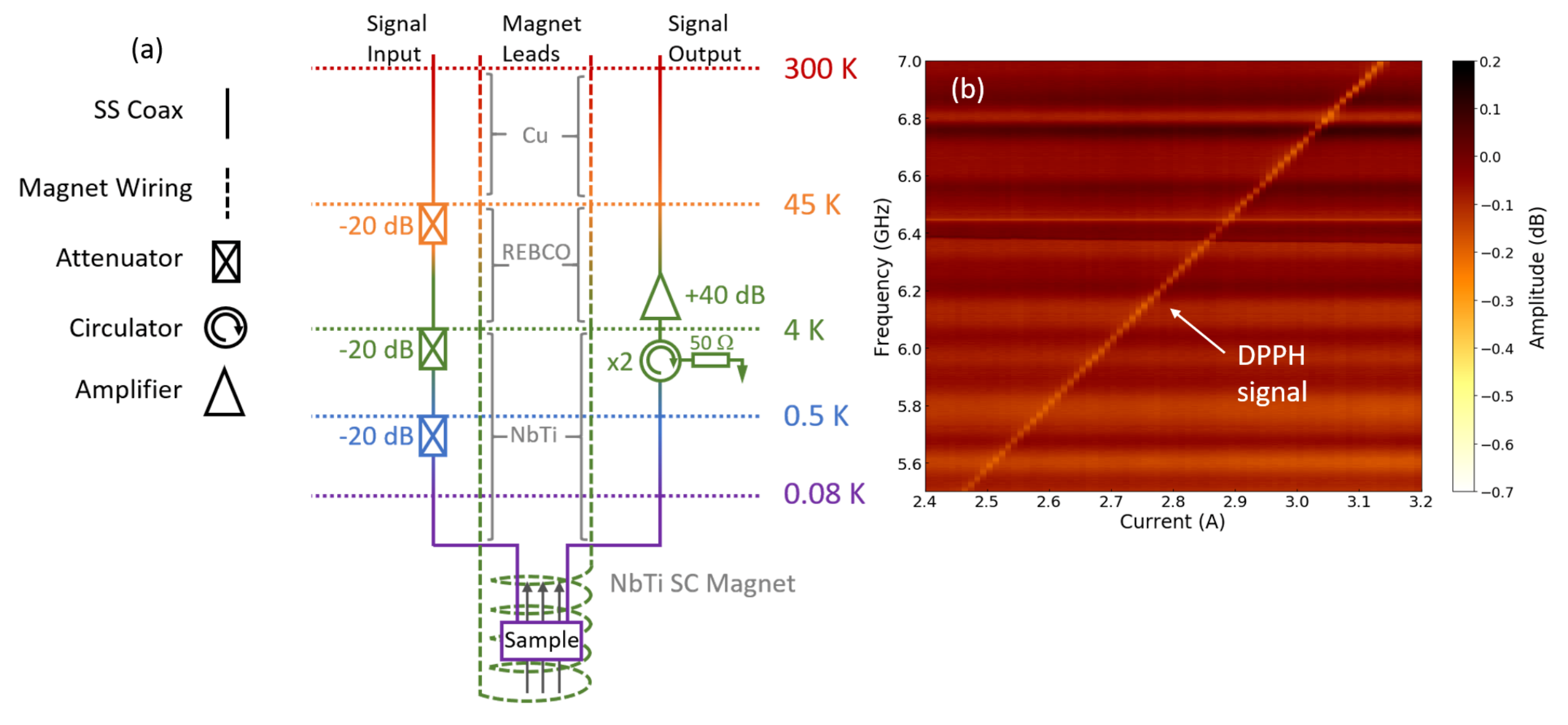}
\caption{(a) Diagram of the adiabatic demagnetization refrigerator experimental setup for microwave transmission measurements. A VNA applied microwave signals to the sample through a series of stainless steel (SS) coaxial cables with 20 dB attenuators mounted on the 45~K, 4~K, and 0.5~K temperature stages. The sample was mounted inside of a NbTi solenoid, oriented such that the NbN transmission line is parallel to solenoid field. Before returning to the VNA, signals pass through a circulator and 40 dB HEMT amplifier. In some measurements an additional 45~dB gain room temperature amplifier was also included. (b) Electron spin resonance measurement of 2,2-diphenyl-1-picrylhydrazyl (DPPH) as a function of solenoid current ($I$). This measurement was used to calibrate the field of the superconducting solenoid. The measurement was performed by placing DPPH on top of a NbN transmission line similar to the one used for the YIG FMR measurements of the main text. From the measurement of the DPPH electron spin resonance signal, we find a field-current calibration of $\mu_0 H = m I + b$ with $m = 78.94 \pm 0.02$~mT/A and an offset of $b=2.32 \pm 0.09$~mT.} \label{adrdiagram}
\end{figure*}

\begin{figure*}
\centering
\includegraphics[scale=0.75]{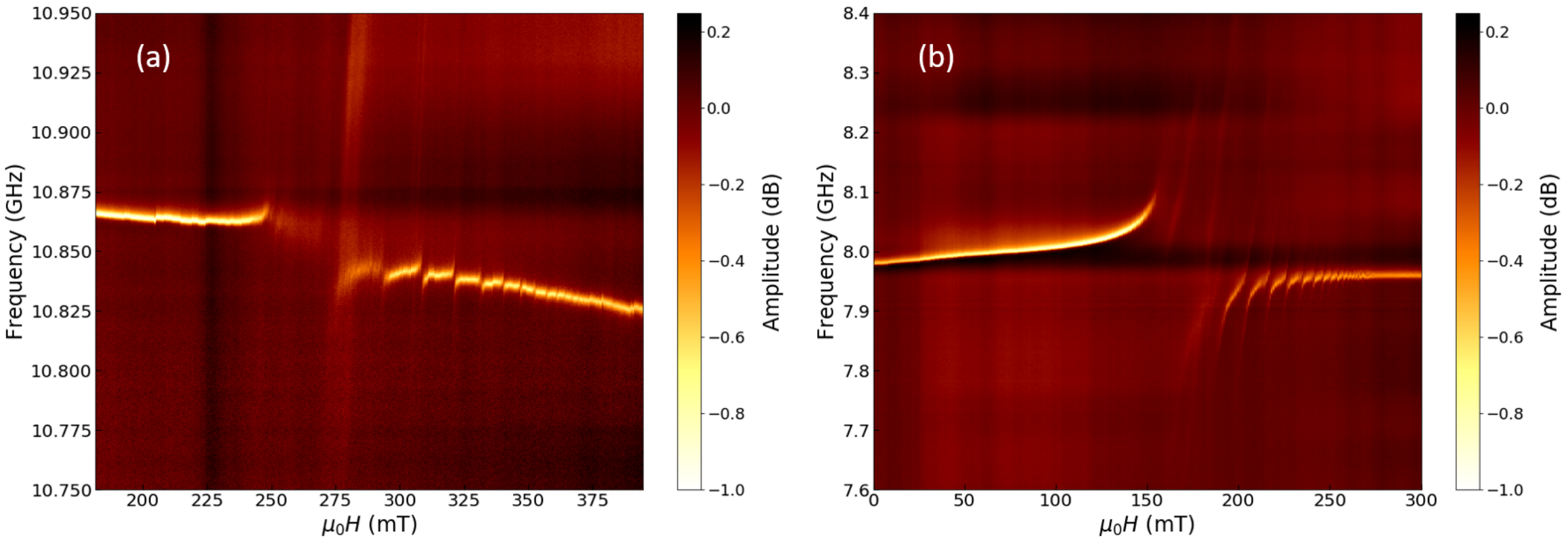}
\caption{Spectra for other resonator devices on the same chip shown in Fig.~1(e) of the main text. These resonances are at (a) 10.866~GHz and (b) 7.980~GHz and show a qualitatively similar behavior as in Fig.~3. As magnetic field is applied, several avoided crossings occur as various spin-wave modes cross the resonator frequency. However, as the YIG sample dimensions are larger, many spin-wave modes (and thus avoided crossings) are clustered together. \label{S2}}
\end{figure*}

\begin{figure*}
\centering
\includegraphics[scale=0.7]{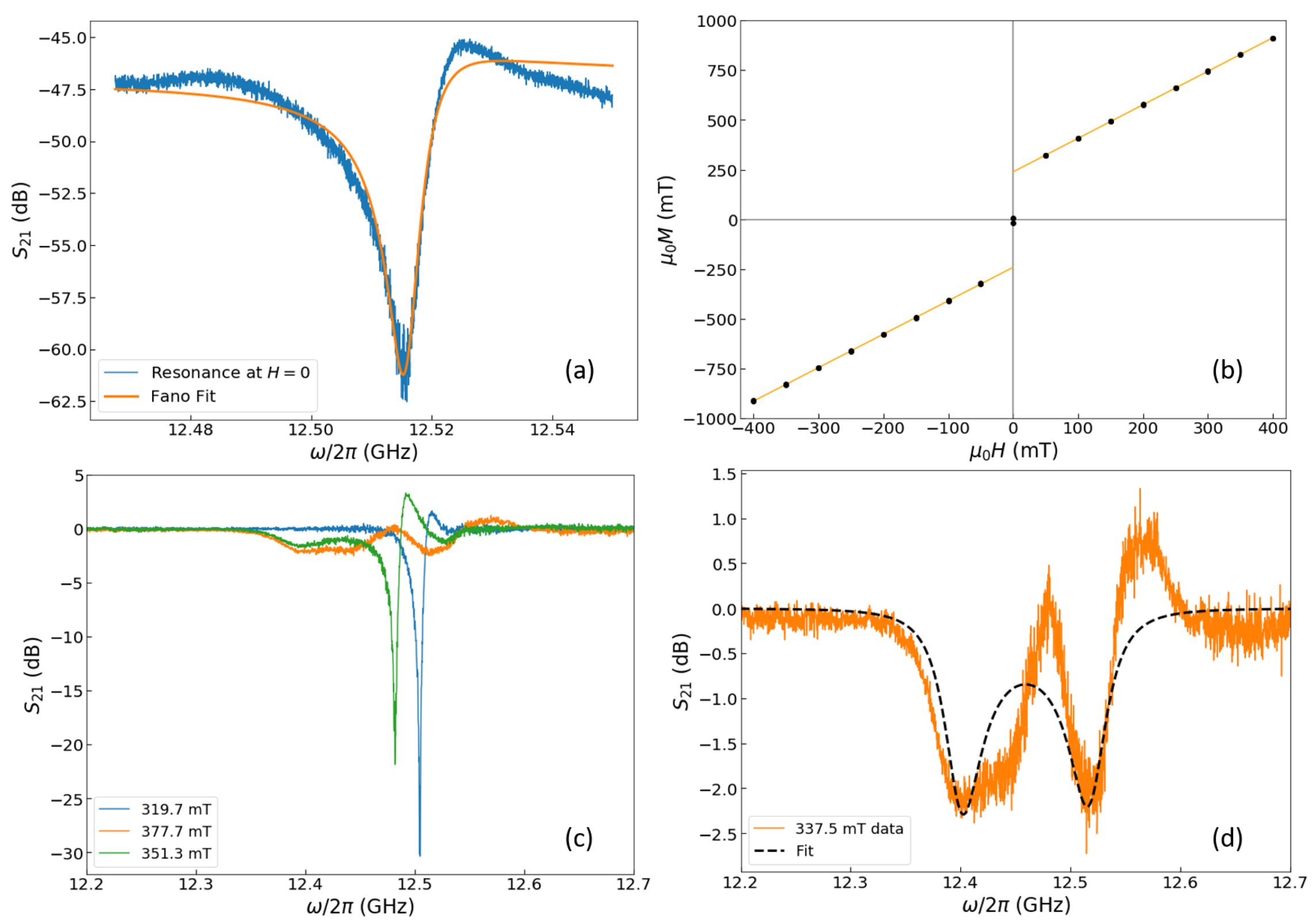}
\caption{(a) Transmission spectrum for a NbN lumped element LC resonator. The resonance is fit using a Fano equation $S_{21} = A(q+\epsilon)^2/(1 + \epsilon^2)$, where $\epsilon = (\omega - \omega_0)/\Gamma$\cite{Fano1961,Limonov2017}, $A$ is the amplitude, $\omega_0$ is the resonance frequency, $\Gamma$ is the linewidth, and $q$ produces the asymmetry of the Fano spectrum. The fit provides a determination of the resonance frequency $\omega_0/2\pi = 12.517$~GHz and full width at half maximum linewidth $\Gamma/2\pi = 9.2$~MHz. (b) Magnetization measurement of a 100 $\mu$m thick film of YIG grown on GGG (black dots). Magnetization was measured at 4~K in a Physical Property Measurement System using SQUID magnetometry. Saturation magnetization is determined to be $\mu_0M_s=240\pm1$~mT after subtracting away the paramagnetic background (orange fits). (c) Transmission $S_{21}$ spectra at different fields for the first avoided crossing shown in Fig.~3 of the main text. The sample spectra shown are at fields before, within, and after the first avoided crossing. (d) Fit of the $\mu_0H = 337.5$~mT data within the avoided crossing. Using Eq.~\ref{S21eq}, we fit the transmission to determine the magnon and resonator linewidths $\kappa = 40 \pm 5$~MHz and $\Gamma = 6.4 \pm 0.8$~MHz, respectively.} \label{S3}
\end{figure*}

Below are supplemental figures for the main text of the manuscript "Strong magnon-photon coupling with chip-integrated YIG in the zero-temperature limit." Figures \ref{adrdiagram}(a) and (b) show the diagram of the cryogenic and measurement setups and calibration data for the magnet, respectively. Figures \ref{S2}(a) and (b) show the spectra of two other devices on the same chip as sample 3. Figure \ref{S3}(a) shows the resonance at zero-field for the lumped-element resonator of sample 3. Figure \ref{S3}(b) shows a SQUID-based magnetization measurement of an unstructured 100~$\mu$m thick film on a GGG substrate. Figures \ref{S3}(c) and (d) show sample transmission spectra $S_{21}$ for the first avoided crossing of Fig.~3 of the main text. The transmission spectrum at $\mu_0H=337.7$~mT is fit to determine the magnon linewidth $\kappa$ as further explained below.

The experiments in the main text were carried out using a superconducting solenoid to apply magnetic fields to the chips. The field strength was calibrated using an electron spin resonance (ESR) measurement of 2,2-diphenyl-1-picrylhydrazyl (DPPH). This measurement was performed by placing powdered DPPH on top of a superconducting CPW transmission line in a similar fashion to a previous study on DPPH ESR\cite{Voesch2015}. The ESR signal was measured over a range from 3.8 to 13.3 GHz and solenoid current 0 to 6 A. The frequency of the spin resonance is described as $f_{DPPH} = \gamma_e\mu_0 H$, where $\gamma_e = 28$~GHz/T is the electron gyromagnetic constant and $H$ is the solenoid field. The field is a function of current ($I$) and turn denisity ($n$): $H = n I$. Therefore, a linear dependence $f_{DPPH} = \gamma_{e} \mu_0 n I$ is expected between frequency of the DPPH ESR signal and the solenoid current. By fitting the frequency of the ESR signal with respect to current, the calibration for magnetic field $\mu_0H = m I + b$ can be extracted. Here, $m = \mu_0 n$ is the calibration constant, and $b$ is an offset field which can arise from residual currents within the solenoid. For our solenoid magnet, we find $m = 78.94 \pm 0.02$~mT/A and $b = 2.32 \pm 0.09$~mT. This is in relatively good agreement with the expected calibration constant $m = 83.3$~mT/A based on the design turn density $n = 66.3$~mm$^{-1}$. For the calibration of the fields in the main text, we do not include an offset field, since such contributions have a history dependence and therefore change over time\cite{Green1971}. Nevertheless, we can estimate that this field contribution only accounts for a $\sim 1\%$ error for the field range of interest and does not strongly affect determinations of saturation magnetization or change the conclusions drawn.

The sample dimensions and their relative errors are listed in Table~\ref{DemagTable}. The errors $\sigma_i$ ($i = x,y,z$) represent the deviations from an ideal cuboid shape. In this case, they are calculated from differences on measurements of opposite edges of each crystal. The demagnetization factors $D_i$ are calculated from equations provided in Ref.~\onlinecite{Aharoni1998}, and the errors $\sigma_{D_i}$ on the demagnetization factors are then calculated as 
$(\sigma_{D_i})^2 = (\frac{dD_i}{dx}\sigma_x)^2 + (\frac{dD_i}{dy}\sigma_y)^2 + (\frac{dD_i}{dz}\sigma_z)^2$.
Since the regression fits determining $M_s$, $g$, $\kappa$, and $\gamma$ do not include errors on the demagnetization factors, the relevant errors are calculated using standard error propagation techniques. For example, the error $\sigma_{M_s}$ on $M_s$ is calculated as
\begin{equation}
    \left(\frac{\sigma_{M_s}}{M_s}\right)^2 = \left(\frac{\sigma_{D_x}}{D_x}\right)^2 + \left(\frac{\sigma_{D_y}}{D_y}\right)^2 + \left(\frac{\sigma_{D_z}}{D_z}\right)^2 +
    \left(\frac{\sigma_H}{H}\right)^2 + \left(\frac{\sigma_{M_{fit}}}{M_{fit}}\right)^2.
\end{equation}
Here, $M_{fit} = M_s$ is the value of the saturation magnetization from the regression, and $\sigma_{M_{fit}}$ is the regression error. As stated above, the field error $\sigma_{H}$ is approximately $1\%$ of the field $H$. Since the demagnetization factors and field are determined independently, covariance terms are not included for error propagation purposes. By this method, the uncertainty values placed on given experimental results take into account errors stemming from imperfect cuboid shapes.

As discussed in the main text, a new approach has been proposed for modeling the magnon-photon coupling $g$ using electromagnetic perturbation theory\cite{RairsPaper}. In these terms, the coupling $g$ can be modeled as
\begin{equation}
g = \sqrt{\frac{\chi \omega_{FMR} \omega_p V_m}{2V_p}},
\end{equation}
where $\omega_{FMR}$ and $\omega_p$ are the ferromagnetic and photon resonance frequencies, respectively, $V_m$ is the volume of the ferromagnet (or ferrimagnet in the case of YIG), and $V_p$ is the mode volume of the resonator. $\chi$ is the YIG susceptibility, which depends on the saturation magnetization $M_s$ as $\chi=\frac{M_s}{(H+H_k+(D_x-D_z)M_s)}$ with magnetocrystalline anisotropy fields $H_k$ and demagnetization factors $D_x$ and $D_z$. Using the $\mu_0M_s=205\pm5$~mT and $\mu_0H=337.5$~mT for the condition that $\omega_{FMR}=\omega_p=12.494$~GHz, we calculate a coupling constant $g=68\pm3$~MHz. This value of $g$ is in close agreement with the value $g=70\pm3$~MHz, which is found using Eq.~3 of the main text and an assumed spin density. Therefore, we can conclude that the effective spin density $\rho$ of our YIG samples does not deviate strongly from the ideal case of $\rho=2\times10^{22}$~cm$^{-3}$ for YIG.

To determine the magnon linewidth $\kappa$, the transmission spectrum $S_{21}$ is modeled as\cite{Morris2017,Boventer2018}
\begin{equation}
S_{21} = 1 - \frac{2\Gamma_c}{i(\omega_p-\omega - \delta) + \Gamma + \frac{g^2}{i(\omega_{FMR}-\omega-\delta)+\kappa}}. \label{S21eq}
\end{equation}
Here, $\Gamma$ is the total linewidth of the bare (i.e.~YIG-free) resonator, and $\Gamma_c$ is the coupling loss rate of the resonator to the transmission line. Eq.~\ref{S21eq} is adjusted to account for the notch-type geometry of the resonator circuit. Figure \ref{S3}(d) shows a fit of the transmission spectrum at $\mu_0H=337.5$~mT within the first avoided crossing of Fig.~3. Like the fits of $\omega_{\pm}$, the resonator frequency is set at the predetermined value $\omega_{p} = 12.494$~GHz. Since $\omega_{FMR}$ depends on the saturation magnetization, $M_s$ is also set as a predetermined constant $\mu_0 M_s = 205$~mT. Likewise, the coupling constant is fixed at $g = 63$~MHz. To accommodate for errors on $\omega_p$ and $M_s$, we include a free parameter $\delta$ that allows for a shift of the horizontal frequency position of the fit. Here, our main focus is the determination of the linewidths, instead of precise determinations of $M_s$, $g$, or $\omega_p$, which are instead determined from fits of $\omega_{\pm}$. The errors on $\kappa$ and $\Gamma$ are determined by including additional error propagation terms for $g$ and $M_s$ as explained above. From the fit in Fig.~\ref{S3}(d), we determine $\Gamma_c = 4.8 \pm 0.6$~MHz, $\Gamma = 6.4 \pm 0.8$~MHz, and $\kappa = 40 \pm 5$~MHz. The fit is shifted downward in frequency by $\delta = 35.7$~MHz.

From this determination of the linewidths, the average number of photons in the resonator $\langle n_p \rangle$ and number of magnons $\langle n_m \rangle$ in the coupled system can be calculated. For a bare resonator without YIG, the number of photons in the resonator can be expressed as\cite{Schneider2020}
\begin{equation}
    \langle n_p^{bare} \rangle = \frac{4 P_{in}}{\hbar\omega_p^2}\frac{(\omega_p/\Gamma)^2}{(\omega_p/\Gamma_c)}.
\end{equation}
When the YIG FMR signal is tuned to match the resonator frequency, resonator photons excite magnons in the system, and energy is split evenly between the FMR and resonator systems. Therefore, by omitting higher order terms (e.g.~two-magnon processes) and assuming a one-to-one photon to magnon conversion, particle numbers can be estimated as $\langle n_p \rangle = \langle n_m \rangle = 0.5\langle n_p^{bare} \rangle$. For the measurement of the avoided crossing, the estimated power on the chip is -97 dBm and $\langle n_p^{bare} \rangle = (1.2 \pm 0.3) \times 10^4$. Therefore, the average number of magnons and photons is $\langle n_p \rangle = \langle n_m \rangle = (6 \pm 2) \times 10^3.$

%\bibliography{bib}

\end{document}